\documentclass[11pt,a4paper]{article}
\usepackage{amsmath,amsfonts,amssymb}
\usepackage{graphicx}
\usepackage{booktabs}
\usepackage{makecell}
\usepackage{natbib}
\usepackage{hyperref}
\usepackage{geometry}
\usepackage{fancyhdr}
\usepackage{setspace}
\usepackage[labelfont=bf]{caption}
\usepackage{orcidlink}
\usepackage{authblk}
\usepackage{xcolor}
\usepackage{url}

\hypersetup{colorlinks=true, allcolors=blue!50!black}

\title{Cutting Bus Service Without Cutting Access:\\ Persistent Homology of Service Deserts and Timetable Reductions}

\author[1,*]{Fabiana Ferracina \orcidlink{0000-0003-4634-4880}}
\author[2]{Ayane Nakamura\orcidlink{0009-0005-1744-0066}}
\author[3]{Bala Krishnamoorthy \orcidlink{0000-0002-2727-6547}}
\author[1]{Hiroyasu Ando \orcidlink{0000-0003-1102-2291}}

\affil[1]{Mathematics Center for Co-creative Society, Tohoku University, 2-1-1, Katahira, Aoba-ku, Sendai, 980-8577, Japan}
\affil[2]{Department of Industrial and Systems Engineering, Keio University, 3-14-1 Hiyoshi, Kohoku-ku, Yokohama, Kanagawa 223-8522, Japan}
\affil[3]{Department of Mathematics and Statistics, Washington State University, 14204 NE Salmon Creek Avenue, Vancouver, 98686, Washington, USA}

\begin{document}
\maketitle

\begin{abstract}
Japanese bus operators are withdrawing service for lack of drivers, and the choice of which trips to withdraw determines who keeps a bus. We measure service at each stop from vehicle GPS and smart-card records, without timetables, and use persistent homology of walking and walking-plus-waiting time to find enclosed service deserts: pockets surrounded by service but beyond walking distance from it. Around Tsukuba, 35 such pockets lie more than 800m from any station or hourly stop. The 17 that are fully ringed by walkable service hold 18,446 residents. We compare two 2020 timetable reductions by Kanto Railway with random reductions and a heuristic frontier of same-size, capacity-feasible alternatives. At two and four buses an hour, both reductions protected service better than random reductions. No alternative found by the sweep dominates the state-of-emergency timetable. The summer timetable is dominated: it withdrew hourly service from 5,253 residents around stops with almost no morning riders, whereas same-size alternatives kept it for all of them at equal waiting. Averages hide a tail: under the actual timetables, 5.9\% and 3.9\% of residents are at stops whose expected wait rises by 10 minutes or more, against 0.5\% and 0.3\% in the alternatives, and capping the increase at 10 minutes at every served stop is feasible at the operator's sizes. If half the removed hours become paid hours saved, the labor saving of both reductions and their main alternatives fall short of the fare revenue on the removed trips.
\end{abstract}


\medskip
\noindent\textbf{Highlights}
\begin{itemize}\itemsep0pt
\item Persistent homology locates enclosed bus service deserts from GPS and smart-card data
\item Access topology varies more between peak and midday than across two years
\item Two 2020 service reductions protected frequent service far better than chance
\item A same-size summer reduction could have kept hourly service for every resident
\item A 10-minute cap on added waiting at every stop is feasible at the same sizes
\end{itemize}

\section{Introduction}

Japanese bus operators are reducing service on a wide scale, and the main reason is usually a shortage of drivers rather than a shortage of riders. In a 2023 survey of 127 major route-bus operators, 77\% had cut or abandoned services or planned to do so within that year, and more than 80\% when plans for the following year are included \citep{tdb2023bus}. The Nihon Bus Association projects that by 2030 operators will have about 93,000 drivers against about 129,000 needed to maintain current routes, a shortfall of 36,000 \citep{nba2023hearing}, and the limits on the working hours of drivers that were introduced in April 2024 have tightened the constraint further \citep{mlitkanto2025driver}. In the Kanto region, 65.6\% of the municipalities with route-bus service experienced route-bus reductions between April 2021 and December 2024, including 70.3\% of the municipalities in Ibaraki Prefecture, and the driver shortage was cited in 85.4\% of the cases reported from April 2024 onward \citep{mlitkanto2025driver}. Kanto Railway, the main bus operator in and around Tsukuba, cut 235 weekday trips (8.5\%) in eight municipalities on December 20, 2023 \citep{kantetsu2023genbin,newstsukuba2023cuts}. At that time it had 533 drivers, against a stated need of 600 \citep{newstsukuba2023drivers}. Further route abolitions were announced for 2026 \citep{kantetsu2025haishi}.

When the scarce resource is driver time, the decision an operator faces is not whether to reduce service but which bus-hours to give up, and this decision determines who continues to have a bus. The two extreme cases differ in what they leave residents. If a corridor is thinned from eight buses an hour to seven, waiting times become longer, but every stop is still served. If the only morning trip on a rural branch is removed, the people who live along that branch are left with no bus. Transit planning has long described this situation as a trade-off between ridership and coverage \citep{walker2012human,murray2003tradeoffs,giuffrida2021dilemma}, and frequency-setting models allocate a fixed fleet across routes so as to balance the two \citep{furth1981setting,ceder2016public,ibarrarojas2015planning,verbas2015tradeoffs}. Nevertheless, reductions are still reported mainly as the number of trips removed. The number of residents who are left without adequate service, and whether a reduction of the same size but with a different shape would have left fewer of them, are seldom measured.

Measuring these quantities requires a description of where the access gaps lie, and data on the service that was actually operated. Regarding the former, it is useful to distinguish between two kinds of gap. A district at the edge of a network may be poorly served, but an extension of the nearest route would reach it. A district that is surrounded by service on every side, but lies beyond walking distance of all of it, is a different kind of gap. It is a hole rather than an edge, and it can be closed only by a route that crosses it, not by a route that is extended toward it. Coverage ratios, distance-to-stop maps and the transit-desert indices built on them \citep{jiao2013transit,jiao2017texas,allen2019sizing} evaluate access location by location, and they do not separate these two cases. Persistent homology makes this distinction directly \citep{carlsson2009,ghrist2008,edelsbrunner2010computational}. As the walking catchments around served stops grow, a one-dimensional homology class records a loop of service, and the death time of the class measures how far the interior of that loop lies from the nearest stop. This idea has been used to locate coverage holes in sensor networks \citep{desilva2007coverage}, and to study the placement of polling sites \citep{hickok2024} and of cooling centers \citep{oneil2026cooling}. \citet{hickok2024} delay the catchment of each site by its waiting time, so that their filtration measures the total time needed to vote. We adapt the same construction to bus service, and we take the waiting time at each stop from the observed frequency of buses.

The other part of the problem is about data. Most accessibility tools assume that a GTFS timetable is available, and Kanto Railway does not publish one. So we measure service from the movements of the buses themselves. We use fleet positions from GPS, restricted to the roads used by revenue trips, together with the smart-card (IC) records of the operator, which show where passengers boarded \citep{pelletier2011smart}. Earlier studies have built GTFS feeds from GPS traces, including for a semi-formal bus system that had no published timetable \citep{williams2015matatu,mansurova2025gps}. In this study we use the traces to rate the service at every stop directly, and we check the resulting stop set against OpenStreetMap and against raw GPS evidence that buses stopped there.

We apply the method to two reductions made by Kanto Railway in 2020. During the COVID-19 state of emergency in Ibaraki, from April 16 to May 14, 2020 \citep{nippon2020emergency}, the operator reduced morning service by 17.7\% of bus-hours, while card boardings fell to about one third of their July level. The legal period is that of the declaration, but the service measurements use the weekdays from April 20 to May 13, 2020. In August 2020, the summer school-holiday timetable reduced morning service by 12.5\%. This second reduction was a planned change that recurs every year, not a response to an emergency. For each reduction, we count the residents who lost access at one, two and four buses an hour, and we compare the result with random reductions of the same size. We also trace a heuristic trade-off frontier between the number of residents losing access and the added waiting of riders, among alternatives of the same size that respect vehicle capacity.

The paper makes four contributions. First, we construct an exact topological map of access that separates enclosed gaps from peripheral ones. The map is computed with an alpha complex for walking distance and with a cubical complex for walking-plus-waiting time, and it is checked against GUDHI. Second, we measure stop-level service without any timetable, from GPS and IC data alone, and we show that the resulting service deserts are not artifacts of missing stops or of service provided by other operators. Third, we use distances between persistence diagrams to measure how much the access map changes over the course of a day, over two years of network change, and under an actual service reduction. Fourth, we compare two actual reductions with chance and with a heuristic trade-off frontier of capacity-feasible alternatives, which is traced by a greedy sweep and is not certified as globally optimal. At two and four buses an hour, both reductions protected service better than any random reduction of the same size. Their positions relative to the frontier, however, are different. No alternative found by the sweep dominates the state-of-emergency timetable, while several alternatives dominate the summer timetable. The stop-level analysis explains this difference: the summer reduction withdrew hourly service where few people rode but where as many people lived as elsewhere.

Averages can hide who bears a reduction, and an operator also has to pay for the service it keeps. We therefore ask two further questions of the same plans. First, how are the added waits distributed across stops, boardings and residents, and how much service can still be removed if the burden at every protected stop is limited, either by a maximum increase in expected wait or by a minimum service standard? Second, for which plans do the driver labor costs saved exceed the fare revenue lost, and how large a loss of passengers would remove the gain? Both questions use the same trip units, stop rates, riders and capacity rule as the frontier, so their results can be compared with it directly (Sections~\ref{sec:equity}--\ref{sec:finance} and \ref{sec:res-equity}--\ref{sec:res-sens}).

This study continues our earlier work on the buses of Tsukuba, in which the bus frequency and vehicle fleet of a park-and-ride system were determined from travel-survey demand under an assumed level of service \citep{nakamura2025parkandride}. In that survey, 95\% of the trips into, within and out of Tsukuba that were made by car or bus were made by car. In the present study, we start instead from the service that was actually operated, because this is what matters most to residents who have no car to rely on.

\section{Background}\label{sec:background}

\subsection{Coverage, waiting time and service reduction}

Access to bus service is usually measured by the walking distance to a stop. The Transit Capacity and Quality of Service Manual treats the area within 400~m of a local bus stop, and within 800~m of a rail or BRT station, as the coverage area \citep{tcqsm2013}. Observed access walks are often longer than this. In Montreal, for example, the 85th-percentile distances are about 524~m to buses and 1,259~m to commuter rail \citep{elgeneidy2014walking}. Waiting time is commonly approximated by half the headway, which is appropriate for passengers who arrive at random. For headways above about 11 minutes, passengers increasingly time their arrival to the schedule \citep{fan2009wait}, and timed arrivals also occur at shorter headways \citep{luethi2007arrival,ingvardson2018arrival}.

Operators balance coverage against ridership \citep{walker2012human}. \citet{murray2003tradeoffs} formalized the tension between access and service efficiency, and \citet{giuffrida2021dilemma} studied it for small cities. Network design and frequency-setting models allocate a fleet across routes \citep{kepaptsoglou2009transit,ibarrarojas2015planning,verbas2015tradeoffs}, covering-location models maximize the population within reach of facilities \citep{church1974maximal}, and greedy algorithms have approximation guarantees when the objective is submodular \citep{nemhauser1978analysis}. Closer to our question, \citet{rumpf2021equitable} reallocate an existing fleet to improve equitable access, and \citet{nishiuchi2018tram} used smart-card data to evaluate reductions in rural tram service in Japan. During the pandemic, many operators adjusted their frequencies to the collapse in demand \citep{gkiotsalitis2021covid}. In Japan, route-bus ridership in April and May 2020 was about half of its level in the previous year \citep{mlit2021whitepaper}. Equity in transit supply has been assessed with Lorenz curves and Gini coefficients across areas \citep{delbosc2011lorenz} and with gaps in supply relative to social need \citep{currie2010gaps}; we apply the same idea to the burden of a reduction and not to the level of supply. We are not aware of work that places an executed reduction on a frontier of alternatives of the same size, evaluated in terms of access and waiting.

\subsection{Persistent homology and coverage}

Consider a finite set of points $P$ in the plane, and the union of discs of radius $r$ centered at these points. As $r$ increases, loops form in the union and are later filled. For each loop, persistent homology records the radius $b$ at which the loop forms (its birth) and the radius $d$ at which the region it encloses is filled (its death) \citep{edelsbrunner2010computational}. The alpha complex is a subcomplex of the Delaunay triangulation of $P$, and it has the homotopy type of the union of discs at every radius \citep{edelsbrunner1983shape,edelsbrunner1994alpha,edelsbrunner1995union}. It therefore computes these pairs exactly. Persistence diagrams are stable: if two functions differ by at most $\varepsilon$ everywhere, then the bottleneck distance between their diagrams is at most $\varepsilon$ \citep{cohensteiner2007stability}. For the distance functions of two point sets, $\varepsilon$ is the Hausdorff distance between the sets, so displacing stops by a few tens of meters moves the diagram by at most that amount.

\citet{desilva2007coverage} used homology to certify coverage by sensor networks. \citet{feng2021voting} analyzed geospatial voting data with persistent homology, and \citet{hickok2024} used a weighted Vietoris--Rips filtration on polling sites. In that filtration, the ball around each site begins to grow only after the mean waiting time at the site, and edges are weighted by the travel time between sites. \citet{oneil2026cooling} applied a filtered witness complex to the coverage of cooling centers. Topological studies of public transport have mostly analyzed network graphs, and to our knowledge, coverage gaps in bus service have not been studied with persistent homology.

\section{Study Area and Data}\label{sec:data}

\begin{figure}[!t]
\centering
\includegraphics[width=\textwidth]{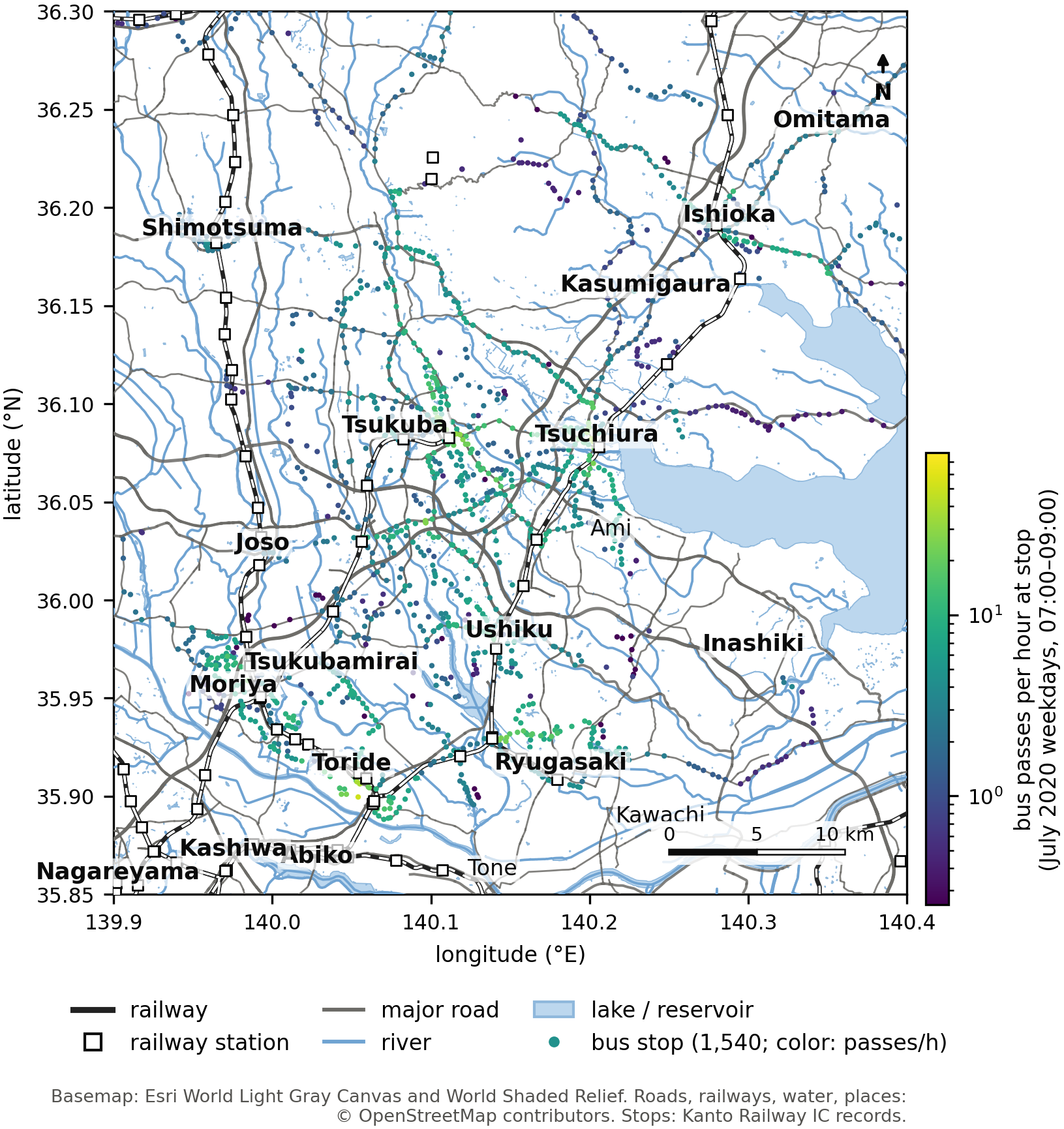}
\caption{The study area. Bus stops are the 1,540 stops located from the IC records, colored by bus passes per hour in the July 2020 weekday morning peak (07:00--09:00, logarithmic scale). Railways, stations, major roads, rivers, lakes and place names are from OpenStreetMap; the basemap is Esri World Light Gray Canvas with World Shaded Relief.}
\label{fig:studyarea}
\end{figure}

Tsukuba is a planned science city in southern Ibaraki Prefecture with about 245,000 residents. It is linked to Tokyo by the Tsukuba Express and is heavily dependent on cars \citep{nakamura2025parkandride}. The study area is the rectangle $35.85$--$36.30^\circ$N, $139.90$--$140.40^\circ$E. It also contains Tsuchiura, Ushiku, Ryugasaki, Moriya and J\=os\=o, and it has 1,784,532 residents in total. All quantities are computed on a 100~m grid. Figure~\ref{fig:studyarea} shows the area with its railways, major roads and rivers, Lake Kasumigaura on its eastern edge, and the bus stops used throughout the paper.

\paragraph{GPS.} Kanto Railway provided daily vehicle-location files for 2020--2022. These files record the position, time, speed, heading and route code of each vehicle at intervals of about 30 seconds.

\paragraph{IC records.} Smart-card boarding and alighting records for 2020 are matched to stops through a table of 1,540 stops with reliable coordinates. Of these stops, 1,088 registered card taps in the July 2020 morning peak. The IC data are used to locate the stops, to identify which GPS records belong to revenue trips, and to provide the demand used to evaluate waiting and vehicle loads (Section~\ref{sec:alternatives}). Since not every passenger pays by card, IC boardings are divided by an assumed card share, for which we use two values.

\paragraph{Population.} Counts from the 2020 Population Census on the 500~m mesh are distributed evenly over the 100~m cells within each mesh cell.

\paragraph{Rail and OpenStreetMap.} The 65 railway stations in the study area, excluding the cable car on Mt.~Tsukuba, are treated as served at every level. OpenStreetMap bus stops of all operators are used for the check described in Section~\ref{sec:gtfs-free}.

\paragraph{Baseline period.} A comparison of the vehicles in the GPS feed with those registering card taps shows that the feed gained about 86 vehicles between March and April 2020, and that 38 of these vehicles had already been carrying card-paying passengers. In other words, the tracking feed was extended, but the service was not. A baseline taken in February 2020 would record this extension as new service that fills two deserts. We therefore use the weekday mornings (07:00--09:00) of July 2020 as the baseline throughout. This period consists of 21 weekdays after the extension and before the reductions that we study. February 2020 is reported only to show the size of the effect.

\section{Methods}\label{sec:methods}

\begin{figure}[!t]
\centering
\includegraphics[width=\textwidth]{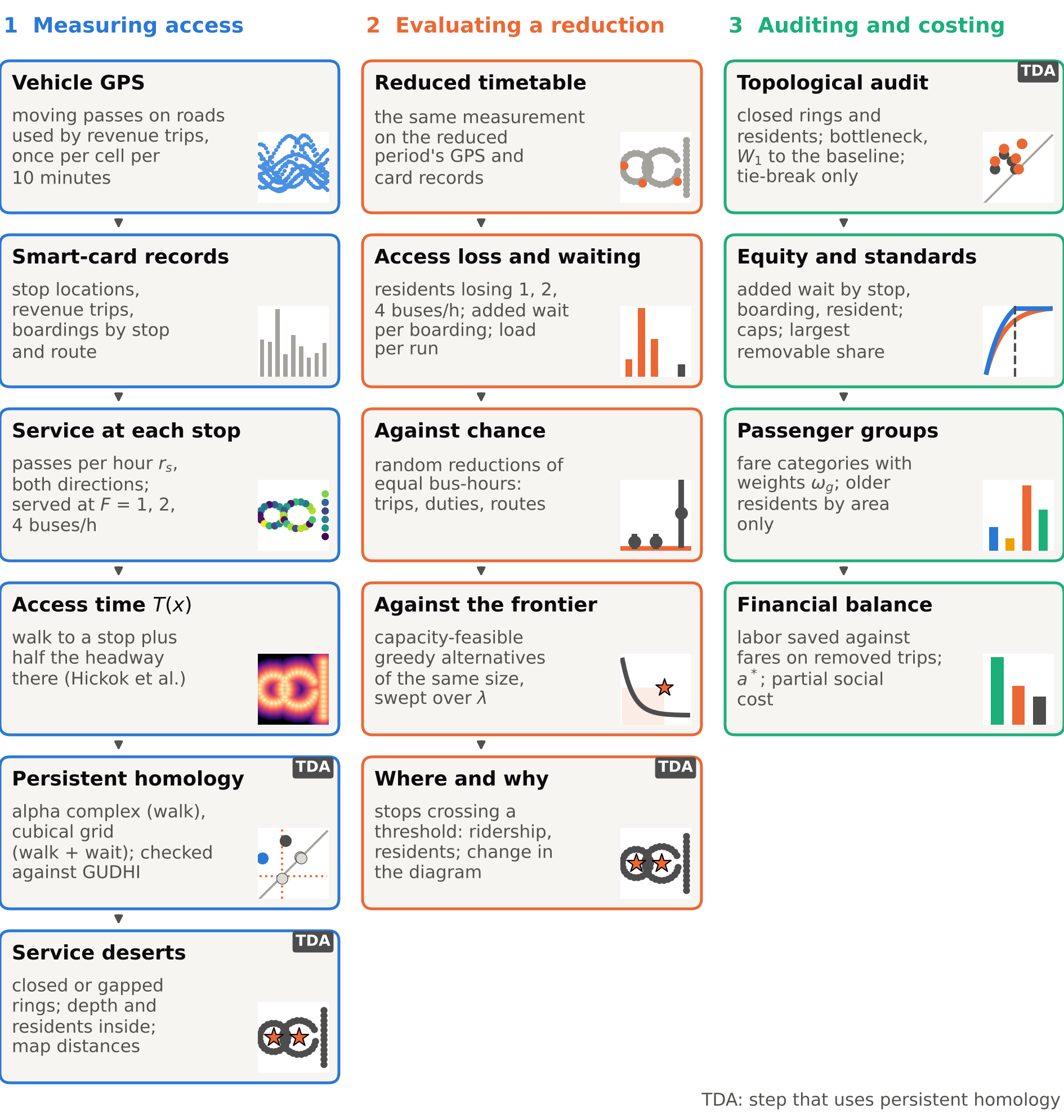}
\caption{Workflow, read from top to bottom in each column. Left column, measuring access: stop-level service is measured from GPS and smart-card records, converted to walking and walking-plus-waiting access, and summarized by persistent homology as enclosed service deserts. Middle column, evaluating a timetable reduction: a reduced timetable is measured in the same way, its access loss and added waiting are computed, and it is compared with random reductions of equal size and with a swept frontier of capacity-feasible alternatives. Right column, auditing and costing every plan: every plan, actual or alternative, is passed through a topological audit and through the equity, passenger-group and financial assessments, with the same trip units, stop rates, riders and capacity rule. The tag TDA marks the stages that use persistent homology (Section~\ref{sec:tda-role}).}
\label{fig:framework}
\end{figure}

Figure~\ref{fig:framework} summarizes the workflow. The left column measures access from the records of the operator. The middle column applies the same measurement to a reduced timetable, and compares the result with random reductions and with a frontier of alternatives. The right column evaluates every plan for its topology, the distribution of its burden, its effect on passenger groups and its financial balance.

\subsection{Service at stops}\label{sec:service}

A GPS record alone does not show that a bus served a stop, because vehicles also idle at depots, run empty between duties and operate express services on highways. We therefore count a \emph{pass} when a vehicle moves at least 15~m between consecutive records through a grid cell, and we count each vehicle once per cell in each 10-minute slot. We keep only the passes on the \emph{in-service footprint}. The footprint consists of the cells crossed at least three times by in-service local trips on sampled days of 2020, dilated by one cell; these trips are reconstructed from card taps, and express and non-revenue route codes are excluded. This definition includes revenue trips on which no passenger tapped a card, and it is our primary measure of offered service. Counting only the trips reconstructed from card taps gives a second, lower measure of service, which we use to confirm the desert map (Section~\ref{sec:res-deserts}).

The service rate $r_s$ at stop $s$ in a time window is the mean number of passes per hour in the busiest cell within one cell of the stop. It is averaged over the weekdays of the window, with both directions combined. A stop is served at level $F$ if $r_s \geq F$, and we report results for $F = 1, 2$ and 4 buses per hour.

\subsection{Access topology}\label{sec:topology}

\paragraph{Walking distance.} Let $P_F$ be the set of stops served at level $F$ together with the railway stations. We compute the $H_1$ persistence diagram of the distance function $d_{P_F}(x) = \min_{p \in P_F} \lVert x - p \rVert$ from the alpha complex of $P_F$, pairing edges with triangles by a union--find sweep under the elder rule. The birth $b$ of a class is the catchment radius at which service first surrounds a region. Its death $d$ is the distance from the deepest point of the region to the nearest served stop. A class with $d > 800$~m and persistence $d - b \geq 300$~m is called an \emph{enclosed service desert}. Its ring is \emph{closed} if $b \leq 800$~m, which means that the 800~m catchments of the stops around it overlap all the way around. Otherwise the ring is \emph{gapped}. The residents of a desert are those in the connected region, beyond 800~m of every served point, that contains the deepest point of the desert. Regions that reach the boundary of the study area are open to the outside and are not counted, and a region that contains several classes is counted only once. We use the generous 800~m catchment for bus stops, so every desert reported here is also a desert under the 400~m standard.

\begin{figure}[!t]
\centering
\includegraphics[width=\textwidth]{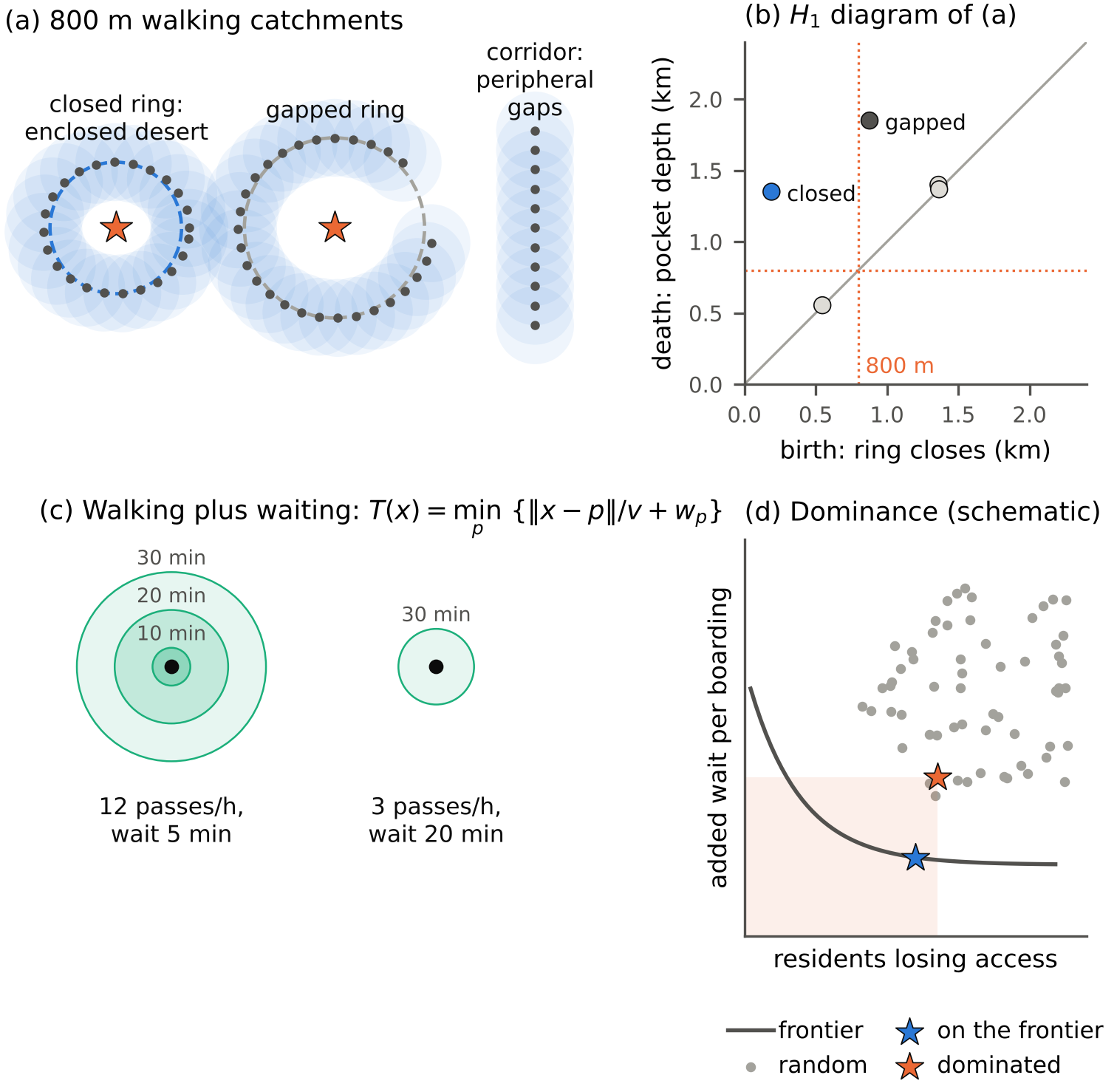}
\caption{Constructions used in the paper, on a small illustrative network. (a) Stops (dots) with 800~m walking catchments; stars mark the deepest point of each loop and dashed circles the largest empty disc, whose radius is the loop's death. (b) The $H_1$ persistence diagram of (a), computed with the alpha-complex code used for the Tsukuba network; blue: closed ring (birth $\leq 800$~m), dark gray: gapped ring, light gray: persistence below 300~m. (c) In the walking-plus-waiting filtration a stop's catchment starts to grow only after its expected wait, so a stop with 12 passes an hour reaches much farther within 20 minutes than a stop with 3. (d) Schematic of the comparison of Section~\ref{sec:alternatives}: a reduction is dominated if an alternative of the same size lies in the shaded region, with no more residents losing access and no more added waiting.}
\label{fig:concepts}
\end{figure}

Figure~\ref{fig:concepts}a--b illustrates these definitions on a small network, computed with the same code. A ring of stops whose catchments close at 190~m encloses a pocket that is 1,351~m deep; this is a closed-ring desert. A ring with a wide gap closes only at 878~m, and it is therefore gapped. A straight corridor forms no loop, and the areas beside it are peripheral gaps, which the diagram does not report as deserts.

\paragraph{Walking and waiting.} To allow frequent and infrequent service to count differently, we use the access time
\begin{equation}
T(x) = \min_{p \in P} \left( \frac{\lVert x - p \rVert}{v} + w_p \right), \qquad w_s = \min\!\left(\frac{60}{r_s}, 60\right)\ \text{min}, \qquad w_{\text{rail}} = \frac{30}{f_{\text{rail}}}\ \text{min},
\label{eq:access}
\end{equation}
with walking speed $v = 80$~m/min and $f_{\text{rail}} = 4$ trains per hour, and with $P$ the set of stops with $r_s \geq 0.25$ per hour together with the stations. Waits are rounded to the nearest minute (8 minutes at stations). This rounding allows $T$ to be computed exactly on the grid with one distance transform per value of the wait (Supplementary Section S4). Since $r_s$ counts both directions, $60/r_s$ equals half the headway in one direction on a two-way street when the two directions have about the same frequency (see the assumptions below). The sublevel set $\{T \leq t\}$ is the union of discs of radius $v(t - w_p)$, each of which begins to grow only once $t$ exceeds the waiting time at $p$. This is the construction of \citet{hickok2024}, with walking in place of multimodal travel and with service frequency in place of queuing. We compute its $H_1$ persistence on the 100~m grid. An \emph{access desert} is a class whose death exceeds $A = 20$ minutes, and its ring is closed if its birth is at most $A$.

\paragraph{Assumptions of the waiting-time model.} The wait $60/r_s$ is half the headway in one direction only if the two directions have about equal frequencies. If a share $\varphi$ of the passes at a stop serves the direction that a passenger needs, the expected wait is $30/(\varphi r_s)$ minutes, which equals $60/r_s$ at $\varphi = 1/2$ and is longer when service is lopsided (Supplementary Section S4). Where the cap does not bind, the baseline and reduced waits are both multiplied by $1/(2\varphi)$, so the ranking of plans at a stop is unchanged, but the absolute minutes are understated when $\varphi < 1/2$. Random arrival is also a better description of frequent service than of infrequent service, where passengers time their arrival \citep{fan2009wait}. The cap at 60 minutes follows our earlier model of Tsukuba, in which a passenger who cannot be served within an hour leaves the system \citep{nakamura2025parkandride}, and it keeps the random-arrival formula from dominating the results at infrequent stops. As a consequence, every stop with $r_s \leq 1$ has the same wait of 60 minutes, so a reduction from one pass an hour to one pass every two hours adds no waiting in the model, and a stop below 0.25 passes an hour is treated as unserved. The loss of service at infrequent stops is therefore carried by the access losses $L_F$ and not by the added waiting, which understates the burden of a reduction there. Section~\ref{sec:res-sens} reports the results for caps of 30, 60, 120 and 240 minutes and for $\varphi$ of 0.5, 0.6 and 0.75.

\paragraph{Computation.} On a planar grid, $H_1$ of a sublevel filtration equals $H_0$ of the superlevel filtration of the complement under 4-connectivity (Alexander duality). A single union--find sweep computes this, together with the deepest point, area and population of each class. On the July 2020 stop set, the alpha-complex diagram agrees with GUDHI's \texttt{AlphaComplex} \citep{maria2014} class for class (37 classes, bottleneck distance 0.00~m). It also agrees with the cubical computation on the 100~m grid to within a bottleneck distance of 179~m. This difference reflects the grid resolution and the way in which the grid treats regions that touch the boundary.

\paragraph{Comparing diagrams.} We compare persistence diagrams using the 1-Wasserstein ($W_1$) and bottleneck distances, both computed exactly by optimal matching. The access-time diagrams keep the classes with persistence of at least 2 minutes. By stability, a bottleneck distance of $\delta$ minutes between two such diagrams implies that access time changed by more than $\delta - 2$ minutes somewhere in the study area; the 2 minutes allow for the short-lived classes that were omitted (Supplementary Section S5). To follow the access map through the day, we also compute the diagram for each half hour of the baseline weekdays. This gives a vineyard, in the sense of persistence over time.

\subsection{Validation of the stop set}\label{sec:gtfs-free}

If the GPS-derived stop set omitted stops served by other operators, or by vehicles that are not included in the GPS feed, some deserts could be artifacts. To check this, we took the 1,895 OpenStreetMap bus stops of any operator that lie more than 150~m from every located Kanto Railway stop, and we identified those at which buses actually stop. For this we used raw, unfiltered GPS, and we counted a stop when a vehicle moved less than 10~m between consecutive records within 40~m of the stop. The stops found in this way were added to the served set, and the diagram was recomputed.

\subsection{Random reductions and the trade-off frontier}\label{sec:alternatives}

We compare each reduction with two benchmarks: random reductions of the same size, and a frontier of planned alternatives of the same size. Each comparison is carried out within a single model, so the actual reduction and its benchmarks are always scored in the same way.

\paragraph{Random reductions.} The first model is the grid model of Section~\ref{sec:service}. Starting from the baseline passes, it removes at random either individual vehicle trips (the passes of a vehicle in one 30-minute block), whole driver duties (the work of a vehicle for the day) or whole route pairs (both directions of a route). The share removed matches the share of moving bus-hours that the actual reduction removed, where moving bus-hours are measured as time in motion between GPS records. Trips and duties are each removed independently with that probability, and route pairs are removed in random order until that share of their passes has been removed (Supplementary Section S6). By this measure, the state-of-emergency timetable removed 22.6\% of morning moving bus-hours and the summer timetable removed 15.0\%. We draw 1,000 random reductions of each kind. As a further reference, we also remove the same share of passes from every service at once.

\paragraph{The frontier model.} The second model works at the level at which a timetable planner can act. A \emph{trip unit} is one route pair in one 30-minute block. Its cost is the number of bus-hours it occupies, where each vehicle is counted once per 10-minute slot whether or not it is moving in that slot. The baseline morning comprises 460 such bus-hours per weekday. Of these, 72\% are accounted for by 835 trip units, and the remaining 28\%, which carry no reliable route label, are held fixed. In this accounting, the state-of-emergency timetable removed 17.7\% of bus-hours and the summer timetable removed 12.5\%. These shares are smaller than the 22.6\% and 15.0\% of moving bus-hours reported above for the grid model. The reason is that the moving measure counts only time in motion between GPS fixes, which gives a smaller total than the full 10-minute presence slots counted here. Both measures describe the same two reductions, and to keep them apart we call the two quantities moving bus-hours and (trip-unit) bus-hours throughout. In this model, the service rate at a stop is the number of distinct vehicle passes per hour within reach of the stop. For the baseline, this model and the grid model agree to within about 2\% on the number of residents served at every level.

\paragraph{Access loss.} $L_F$ is the number of residents whose best stop or station within 800~m is served at level $F$ in the baseline but not after the reduction. A resident whose service drops from four buses an hour to none crosses all three thresholds, and is counted once in each of $L_1$, $L_2$ and $L_4$. The sum is therefore a weighted count of threshold crossings, not a count of unique affected residents. For this reason, we also report $L_1$, $L_2$ and $L_4$ separately.

\paragraph{Waiting and capacity.} Morning boardings at each stop and on each route pair are taken from the IC records of the period of the reduction, and are divided by the card share $\sigma$. The added expected waiting time per boarding is the boarding-weighted increase in $w_s$ of Eq.~\eqref{eq:access}. The added waiting time per resident weights the same increase by the residents for whom the stop is the best stop. A reduction is \emph{capacity-feasible} if no route pair carries more than $C$ boardings per remaining run on average. We compute every frontier twice, with $(C, \sigma) = (50, 0.85)$ and $(60, 0.75)$. The second setting raises estimated ridership by 13\% and allows 20\% more passengers per run.

\paragraph{Frontier.} A lazy greedy search removes, at each step, the trip unit with the smallest increase in
\begin{equation}
L_1 + L_2 + L_4 + \lambda \times (\text{added boarding-minutes of waiting})
\label{eq:objective}
\end{equation}
per bus-hour saved. Only units whose removal keeps the reduction capacity-feasible are considered, and the search continues until the bus-hours removed by the actual reduction are reached. Sweeping $\lambda$ over the values 0, 0.3, 1, 3, 10, 30, 100, 300, 1000 and $10^6$ traces alternatives that range from access-first to waiting-first. An alternative \emph{dominates} the actual reduction if it has no greater access loss and no greater added waiting per boarding, and is strictly better in at least one of the two (Figure~\ref{fig:concepts}d).

\subsection{Role of topology in plan evaluation}\label{sec:tda-role}

The objective in Eq.~\eqref{eq:objective} contains no topological term. Topology therefore does not generate the candidate plans. It enters at two other stages. In \emph{evaluation}, every plan, the actual timetable and each alternative, is passed through the same alpha-complex pipeline as the baseline (Section~\ref{sec:topology}), which gives the number of enclosed deserts, the residents inside closed-ring pockets, and the bottleneck and $W_1$ distances to the baseline diagram. In \emph{selection}, these quantities serve as a tie-breaker: among the plans whose $L_1 + L_2 + L_4$ lies within a tolerance of the smallest value at a given budget, we prefer the plan that opens the fewest closed rings and moves the diagram least. Whether this changes the choice is an empirical question, which Section~\ref{sec:res-topoaudit} answers by asking whether plans that cannot be told apart by $L_F$ and waiting differ in their topology. Adding a topological penalty directly to the objective is possible, but it would require recomputing the alpha complex after every removal, and we leave it to future work. The topological summaries of a plan are defined in Supplementary Section S12.

\subsection{Equity and minimum service standards}\label{sec:equity}

Let $z$ denote a reduction plan, $w_s^0$ the baseline expected wait at stop $s$ (Eq.~\eqref{eq:access}) and $w_s(z)$ its value under the plan, so that $\Delta w_s(z) = w_s(z) - w_s^0$. These are spatial distributions of expected waits under the model, not distributions of the waits that individual passengers realize. We report the distributions of $w_s(z)$ and $\Delta w_s(z)$ over the stops that are served in the baseline, with three explicit weightings that answer different questions: equal weight per stop, weight by morning boardings, and weight by the residents for whom the stop is the best stop in the baseline (the assignment used for the added waiting per resident). For each weighting we give the mean, the median, the 90th, 95th and 99th percentiles, the maximum, the share of weight with any increase and with an increase of at least 10 minutes, the Gini coefficient of the increases \citep{delbosc2011lorenz}, the cumulative distribution, and maps of $\Delta w_s$.

To compare an average-oriented plan with plans that limit local burdens at the same number of bus-hours removed, let $S_{\text{protected}}$ be the set of stops whose service is to be protected. We consider two constraints. The first limits the increase in expected wait,
\begin{equation}
\max_{s \in S_{\text{protected}}} \bigl\{ w_s(z) - w_s^0 \bigr\} \leq \Delta \bar w ,
\label{eq:deltaw}
\end{equation}
and the second is a minimum service standard on the level after the reduction,
\begin{equation}
\max_{s \in S_{\text{protected}}} w_s(z) \leq \bar w .
\label{eq:wbar}
\end{equation}
The second constraint also covers disadvantages that existed before the reduction, and it does not imply that every passenger's realized wait is below $\bar w$. Two cases need a rule. First, a stop whose baseline wait already exceeds $\bar w$ would make the constraint infeasible before any service is removed. We therefore apply Eq.~\eqref{eq:wbar} in the form $w_s(z) \leq \max(\bar w, w_s^0)$, which allows no protected stop to end above the standard or above its baseline wait, whichever is larger. Second, because the wait is capped at 60 minutes, the loss of the last trips at a stop that has one pass an hour or less leaves $w_s$ unchanged. We therefore also require that every protected stop keep at least $\min(r_s^0, r_{\min})$ passes an hour, with $r_{\min}$ a policy parameter (0.5 passes an hour in our runs), so that no protected stop loses its service altogether. The protected set is a policy choice. We compare two: all stops served in the baseline, and the stops with at least 20 morning boardings in the window of the state-of-emergency reduction. The thresholds $\Delta \bar w$ and $\bar w$ are likewise illustrative values that a planning authority would set, and we sweep them.

The constrained plans are produced by the same lazy greedy search with the same objective, restricted to trip units whose removal keeps every protected stop within the constraint. Passes at a stop only decrease as trip units are removed, and waits only increase, so a unit that violates a constraint at one step violates it at every later step, and the lazy evaluation remains valid, exactly as for the capacity rule. If a reduction target cannot be met under a constraint, we report the largest share of bus-hours that can be removed while meeting it, which answers how much service can be removed while the standard is kept. The statistics, the admissibility rule and the proof that the lazy evaluation remains valid are given in Supplementary Section S9.

\subsection{Passenger groups}\label{sec:groups}

The IC records carry four passenger counts for each transaction: adults, adults with disability fares, children, and children with disability fares. They contain no age field, so older passengers cannot be identified directly. For the groups $g$ that the fare categories define, let $B_{sg}$ be the reference number of morning boardings by group $g$ at stop $s$, and $\omega_g$ a policy weight on the burden of that group. The weighted increase in expected waiting time is
\begin{equation}
\sum_{g} \omega_g \sum_{s} B_{sg} \bigl[ w_s(z) - w_s^0 \bigr] .
\label{eq:groupwait}
\end{equation}
We obtain $B_{sg}$ by multiplying the boardings at each stop in the period of the reduction by the share of group $g$ among that stop's IC transactions in the baseline, shrunk toward the pooled share for stops with few transactions. The same $B_{sg}$ are used for all plans, as the expression requires. Eq.~\eqref{eq:groupwait} applies the common stop-level wait to all groups, so it does not capture differences in arrival behavior, and it uses card transactions only. In the search, the boardings in the waiting term of Eq.~\eqref{eq:objective} are replaced by $\sum_g \omega_g B_{sg}$. We compare an equal-weight scenario with scenarios that place a weight greater than one on one group; the weight on children also applies to children with disability fares, and the weight on disability fares to adults and children with disability fares. The weights are alternative policy priorities, not estimates. Because older passengers cannot be identified, we also evaluate the impact on areas with larger older populations: the residents aged 65 and over in each census mesh are assigned to their best stop in the baseline, and the added waiting is weighted by them. This is an area-based assessment and not a measurement of the impact on older bus users. The group shares, the weighted search and the proxy are defined in Supplementary Section S10.

\subsection{Financial balance and a partial social cost}\label{sec:finance}

We restrict the cost analysis to driver labor and hold other costs and fares constant. Let $\Delta H(z)$ be the operated bus-hours removed by plan $z$ (in trip-unit bus-hours), $\rho \in (0,1]$ the share of them that becomes paid driver hours saved, through lower overtime or changes in duty allocation, and $c_h$ the hourly labor cost. Let $R(z)$ be the fare revenue on the removed trip units at the demand of the window of the reduction, and $\theta$ the share of that revenue that is retained because passengers switch to another departure and continue to pay the same fare. The change in profit is
\begin{equation}
\Delta \Pi(z) = c_h \, \rho \, \Delta H(z) - (1 - \theta)\, R(z).
\label{eq:profit}
\end{equation}
An improvement in profit does not imply that the operation is profitable overall. The central quantity is the break-even abandonment share, $a^* = c_h \rho \, \Delta H / R$, which is the share of the revenue on removed trips that can be lost for good before the labor savings no longer compensate for it (a value of at least one means that the plan improves profit even if all of that revenue is lost). The $(1-\rho)\Delta H$ operated hours that do not become paid-hour savings are reported separately as released capacity, not as monetary savings. As the wage benchmark we use JPY 1,906 per hour for bus drivers (general workers in enterprises with 10 or more employees, both sexes) in the 2020 Basic Survey on Wage Structure \citep{mhlw2021wage2020}. It is computed as contractual cash earnings per month times 12 plus annual special cash earnings, divided by 12 times the scheduled and overtime hours worked per month: $(296.9 \times 12 + 714.4)$ thousand yen divided by $12 \times (159 + 28)$ hours. The 2025 figure for route bus drivers, JPY 1,907 per hour \citep{mhlw2025jobtag}, is almost the same. This national figure includes overtime pay and bonuses, is not specific to the operator, and does not represent the full labor cost of the employer, which also pays social insurance contributions and other labor costs. We therefore also report $c_h = $ JPY 2,500 and 3,200, about 1.3 and 1.7 times the benchmark; these values are illustrative and not estimates. We vary $\rho$ from 0.25 to 1.

Revenue is taken from the fare field of the IC taps on each trip unit in the baseline window, divided by the card share $\sigma$ and multiplied by the ratio of boardings in the window of the reduction to those in the baseline, so that every plan is valued at the same reduced demand. The taps of the reduction window cannot be used directly, because they record only the service that still ran and therefore carry no revenue for the trips that the operator removed. Fares are held fixed, and $\theta$ is varied rather than estimated, so no demand-response model is required. Evidence on how demand responds to fares and to service quality \citep{paulley2006demand} could inform a plausible range for $\theta$, but we do not use it. The analysis covers the morning peak (07:00--09:00) of weekdays only, and no figure is annualized. For the operator's own reductions, revenue is attributed to each trip unit in proportion to the share of its bus-hours that the timetable removed. Because the reduced timetables also moved trips between 30-minute blocks and between service with and without a route label, the trip units with a route label lost more bus-hours than the timetables saved overall, and we report a range for the operator's reductions (Section~\ref{sec:res-finance}). To show how revenue and access trade off, we also run the access-first search ($\lambda = 0$) with an added term that charges the revenue on the removed trip units at 0, 0.01 and 0.1 residents per yen; we call these the revenue-aware plans. The revenue by trip unit, the attribution for the operator's own reductions and the definitions of the revenue-aware plans and of the break-even benefit are given in Supplementary Section S11.

To describe the passenger time that accompanies the financial gain, we add a partial social cost. With $W(z)$ the total expected waiting time of the passengers who continue to travel and $v$ the monetary value of an hour of waiting, the change in $SC_{\text{partial}}(z) = c_h H(z) + v\,W(z)$ relative to the baseline is $-c_h\rho\,\Delta H + v\,\Delta W$. Fare payments cancel between passengers and the operator. If some trips are abandoned, the benefit they provided is lost and has to be set against the saving; a fall in total waiting caused by fewer passengers must not be counted as an improvement. We therefore report the net change before the loss of benefits, and the break-even benefit per abandoned trip. We set $v = \kappa c_h$ with $\kappa \in \{0, 0.5, 1\}$ purely for illustration, because empirical values of time differ by purpose and by mode \citep{abrantes2011meta}. An alternative, used in our earlier model of Tsukuba, values an hour of waiting at the hourly productivity of the region as an opportunity cost \citep{nakamura2025parkandride}. The measure excludes changes in walking, in-vehicle and transfer time.

\section{Results}\label{sec:results}

\subsection{Enclosed service deserts}\label{sec:res-deserts}

\begin{figure}[!t]
\centering
\includegraphics[width=0.7\textwidth]{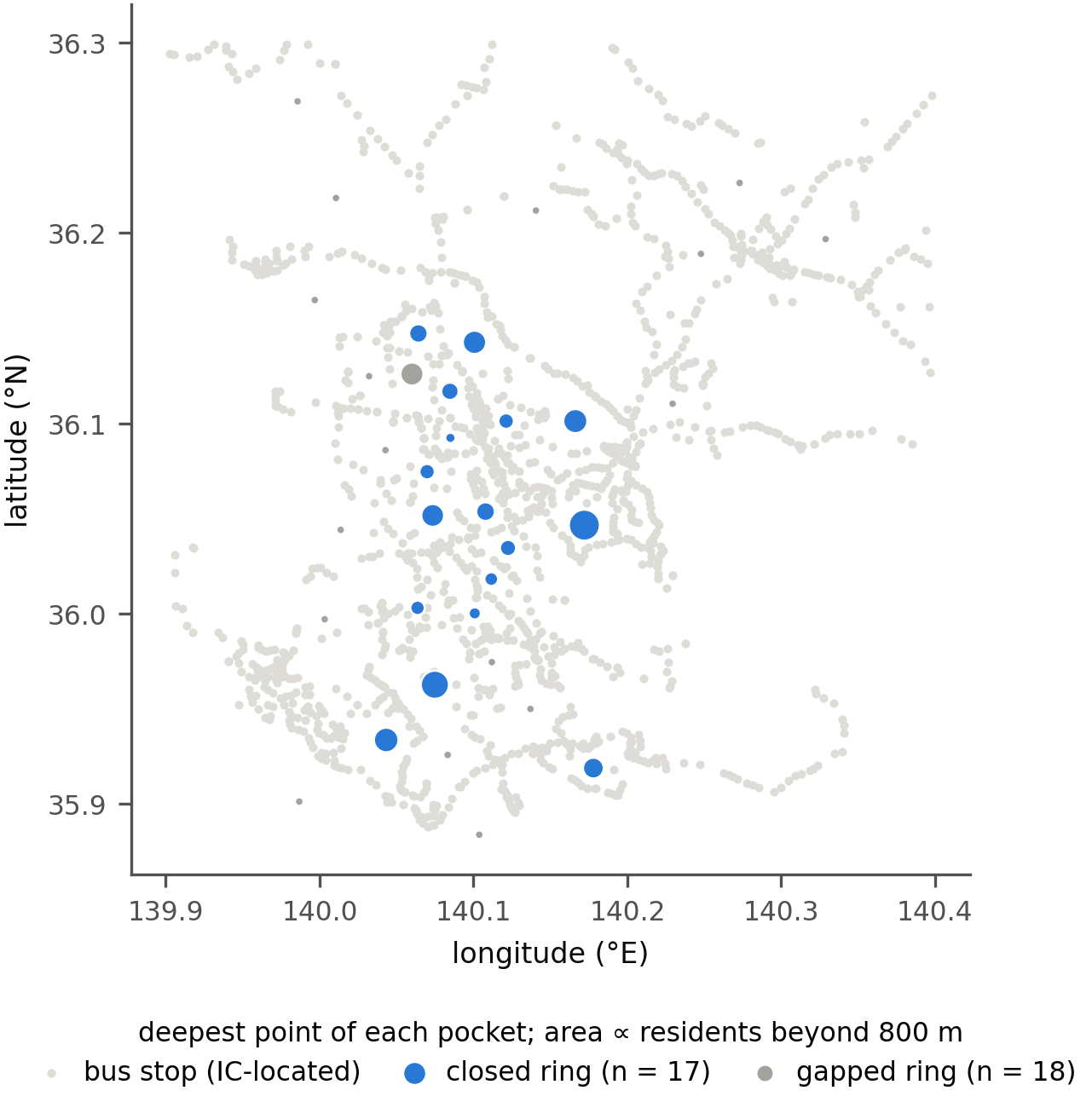}
\caption{Enclosed service deserts in the July 2020 weekday morning peak, for stops with at least one bus an hour. Each marker is the deepest point of an $H_1$ class of the alpha complex with persistence of at least 300~m and death beyond 800~m; blue markers have closed rings (birth $\leq 800$~m) and gray markers gapped rings. Marker area is proportional to the residents beyond walking distance. There are 35 classes, 17 with closed rings, and 19,871 residents, 18,446 of them in closed-ring pockets.}
\label{fig:deserts}
\end{figure}

In the July 2020 morning peak, 1,125,362 of the 1,784,532 residents of the study area live within 800~m of a railway station or of a stop with at least one bus an hour. The corresponding numbers are 1,046,596 for two buses an hour and 935,550 for four. The alpha complex on the 1,407 points served at least hourly has 37 $H_1$ classes with persistence of at least 300~m, and 35 of these classes die beyond 800~m. We call each such class an enclosed service desert (Figure~\ref{fig:deserts}). A class is a topological feature of the filtration. In general, a single geographic pocket of unserved land can contain more than one class, and the interior of one class can be split by later additions to the stop set. The number of classes and the number of geographically distinct pockets therefore need not change together, as the robustness check below illustrates. Here, the 35 classes contain 19,871 residents who are beyond walking distance of any served stop. Seventeen of the classes have closed rings, and these 17 pockets contain 18,446 of those residents. The deserts lie in the interior of the network, in and around central Tsukuba.

\begin{figure}[!t]
\centering
\includegraphics[width=\textwidth]{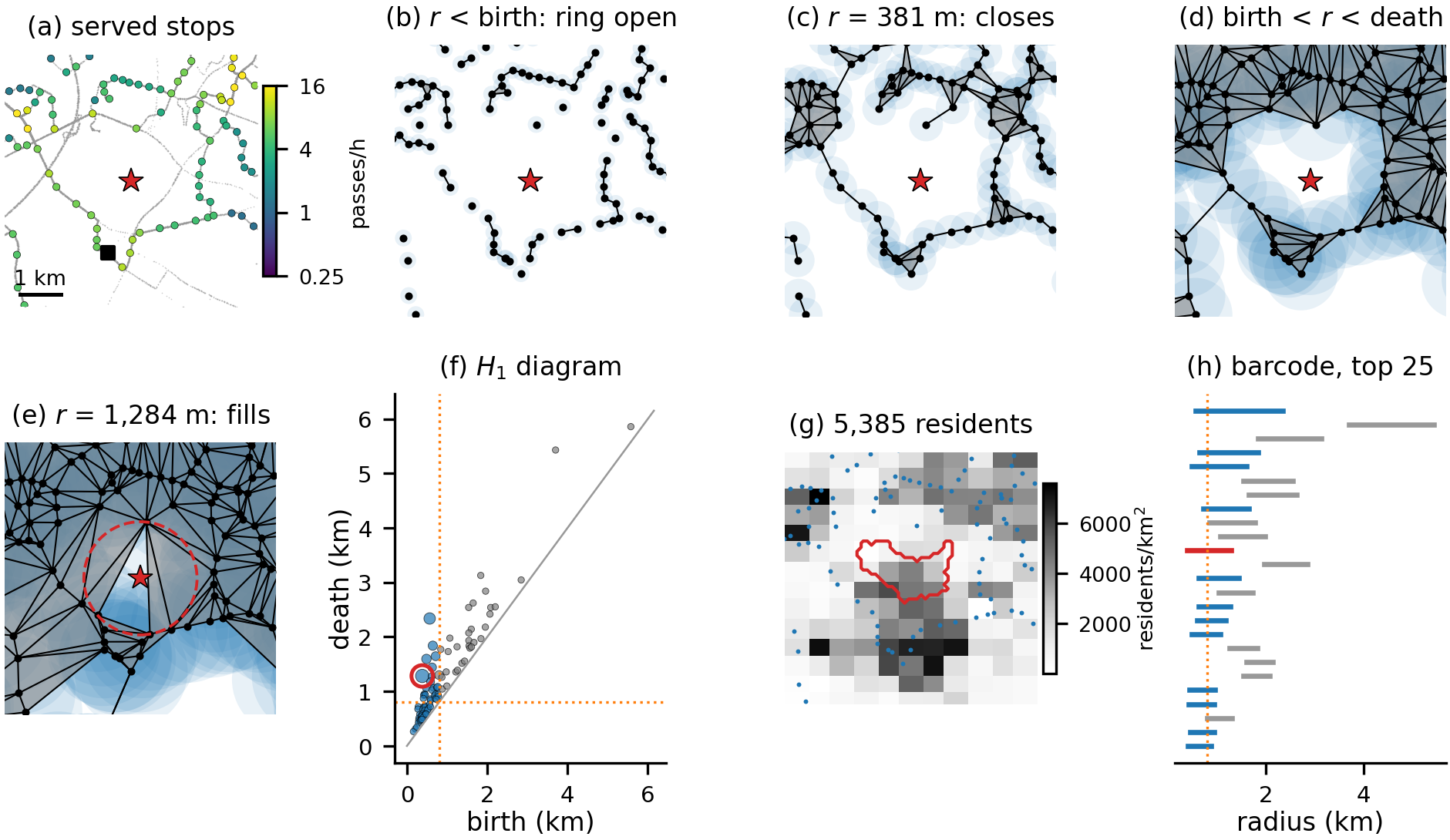}
\caption{From served stops to persistent homology for one enclosed desert. (a) Stops colored by bus passes per hour, with one day of bus GPS positions (gray) and stations (black squares); the star marks the deepest point of the pocket, and the bar is 1~km. (b--e) Alpha complex below the birth radius, at birth (381~m, when the ring closes), between birth and death, and at death (1,284~m, when the pocket fills; dashed circle: the largest empty disc). (f) $H_1$ diagram for stops served at least hourly (blue: closed ring; marker size: residents beyond 800~m; dotted lines: 800~m). (g) Residents per km$^2$ and the area beyond 800~m (red outline). (h) Bars of the 25 most persistent classes, with this pocket in red.}
\label{fig:pipeline}
\end{figure}

Figure~\ref{fig:pipeline} follows the largest of these pockets, centered at 36.046$^\circ$N, 140.172$^\circ$E, from the stops to the diagram. The stops on the routes around it are passed up to 16 times an hour. As the walking catchments grow, the catchments of the stops around the pocket first join into a ring at a radius of 381~m, so no gap between neighboring stops on the ring is wider than 762~m. The region inside is filled only at a radius of 1,284~m. This is the radius of the largest empty disc, whose center is the deepest point of the pocket. The area beyond 800~m contains 5,385 residents. A coverage map would show this area as a patch far from any stop. The persistence diagram shows, in addition, that the area is enclosed and that the ring of service around it closes at a small radius. Because the pocket is enclosed, a short link would serve it. On the lower-bound service map of Section~\ref{sec:service}, where the same pocket contains the same 5,385 residents, a 1.2~km spur from the nearest served road to the deepest point would bring 4,510 of them within 800~m of a stop. Across the ten enclosed deserts where new route length is most effective, 17.5~km of new route would bring 14,775 residents within walking distance.

\begin{figure}[!t]
\centering
\includegraphics[width=\textwidth]{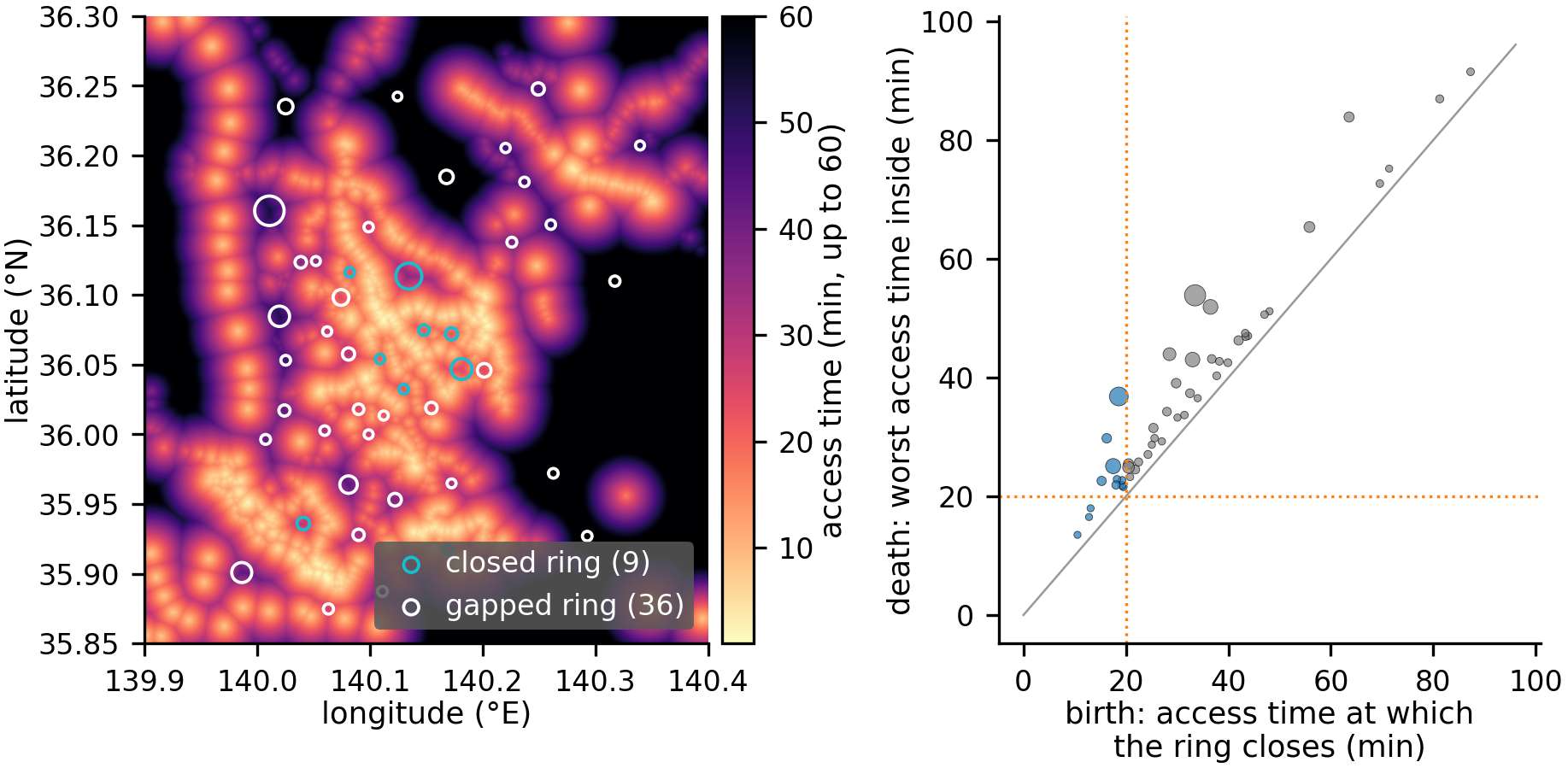}
\caption{Access time (walking plus waiting, Eq.~\ref{eq:access}; shown up to 60 minutes) in the July 2020 weekday morning peak. Circles mark access deserts, regions whose worst access time exceeds 20 minutes: cyan, closed ring; white, gapped ring; circle size, residents. Right: the $H_1$ diagram of the access-time filtration (blue: closed ring; dotted lines: $A = 20$ minutes).}
\label{fig:access}
\end{figure}

Waiting time changes which gaps matter. In the access-time filtration of Eq.~\eqref{eq:access} (Figure~\ref{fig:access}), 45 regions have a worst access time above 20 minutes. Nine of these regions are closed rings, which means that they are surrounded by stops whose 20-minute access areas all overlap. The access deserts contain 49,668 residents, and 14,526 of them are inside closed rings.

\paragraph{Robustness of the desert map.} We carried out three checks to confirm that the deserts are a feature of the network rather than of the data. First, the deserts are not filled by the services of other operators. Of the 4,537 OpenStreetMap bus stops in the study area, 1,895 lie more than 150~m from any located Kanto Railway stop. Raw GPS records buses passing 632 of these stops at least once every four hours, but buses stop at only 59 of them, and only 10 of these 59 are not already counted. Adding these 10 stops changes the number of desert classes deeper than walking distance from 35 to 36, while the number of residents in them remains 19,871. The 36th class is one of the open, gapped deserts that continue past the edge of the study area, and it is not an enclosed pocket, since the number of closed-ring classes and the number of residents inside them remain 17 and 18,446. The new class is also small enough that it does not change the resident total at this precision. This case shows that the class count and the affected population are two different summaries of the same diagram, and that they need not change together. None of the 17 closed-ring pockets contains uncounted service of even one bus an hour, and 12 of them contain no OpenStreetMap bus stop of any operator. Second, the map persists when service is counted only on trips reconstructed from card taps. On the grid, this lower measure gives 26 enclosed deserts beyond 800~m at one bus an hour, 12 of them with closed rings. Because fewer stops count as served under this measure, the pockets are larger, and they contain 42,550 residents. Third, the map does not depend on the implementation: the alpha complex, GUDHI and the cubical grid computation agree, as described in Section~\ref{sec:topology}.

\subsection{Temporal variation of access}\label{sec:res-dynamics}

\begin{figure}[!t]
\centering
\includegraphics[width=0.9\textwidth]{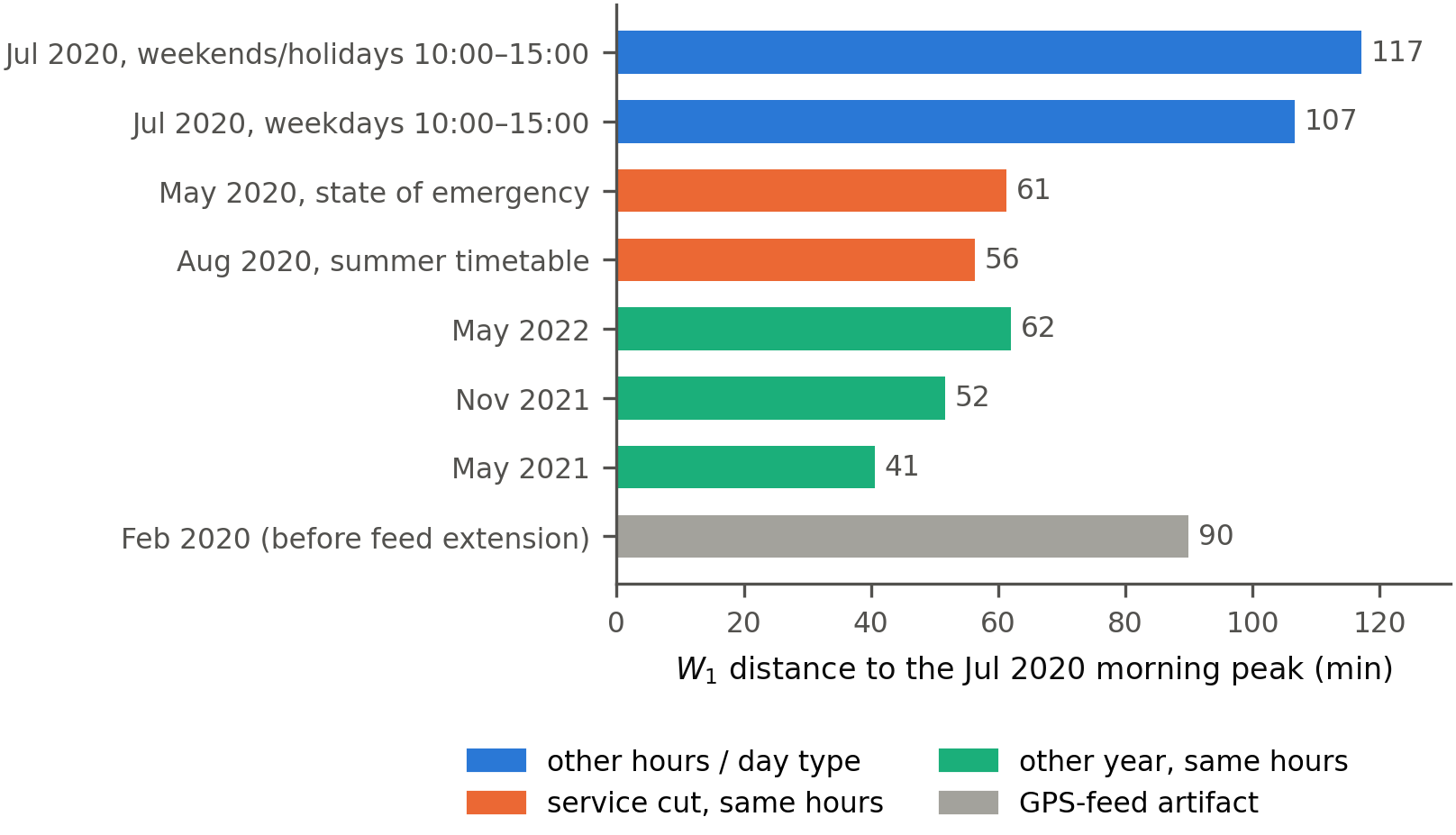}
\caption{Wasserstein distance ($W_1$, minutes) between the access-time persistence diagram of each period and that of the July 2020 weekday morning peak, grouped by what differs from the baseline. Periods are weekday mornings (07:00--09:00) unless stated otherwise.}
\label{fig:windows}
\end{figure}

Figure~\ref{fig:windows} gives the Wasserstein distance between the access-time diagram of each period and that of the July 2020 morning peak. The same July weekdays between 10:00 and 15:00 lie 106.6 minutes from the morning peak, and July weekends and holidays lie 117.1 minutes from it. The morning peaks of May 2021, November 2021 and May 2022 lie 40.6, 51.6 and 62.0 minutes away, and the two reduced timetables lie 61.3 minutes (state of emergency) and 56.3 minutes (summer) away. The time of day therefore changes the topology of access about twice as much as two years of change in the network, and the state-of-emergency timetable changed it by about as much as those two years. February 2020 lies 89.9 minutes away, and this distance reflects the extension of the GPS feed described in Section~\ref{sec:data}.

The change also has a topological form. In the morning peak, nine access deserts are enclosed by closed rings of service. Between 10:00 and 15:00 on July weekdays only one closed ring remains, and on weekends and holidays there are none. As frequencies fall, the rings of service around these deserts no longer close within 20 minutes. The number of residents in access deserts increases by 15\%, from 49,668 in the morning peak to 57,083 at weekday midday, while the number of closed rings falls from nine to one. The diagram distance registers both of these changes, whereas a count of residents registers only the first. The two reductions changed the map in different ways. Under the state-of-emergency timetable, the nine closed rings remained, but the number of residents inside them rose from 14,526 to 24,011, an increase of 65\%. Under the summer timetable, the nine closed rings contained 11,514 residents, which is 21\% fewer than in July.

\begin{figure}[!t]
\centering
\includegraphics[width=0.86\textwidth]{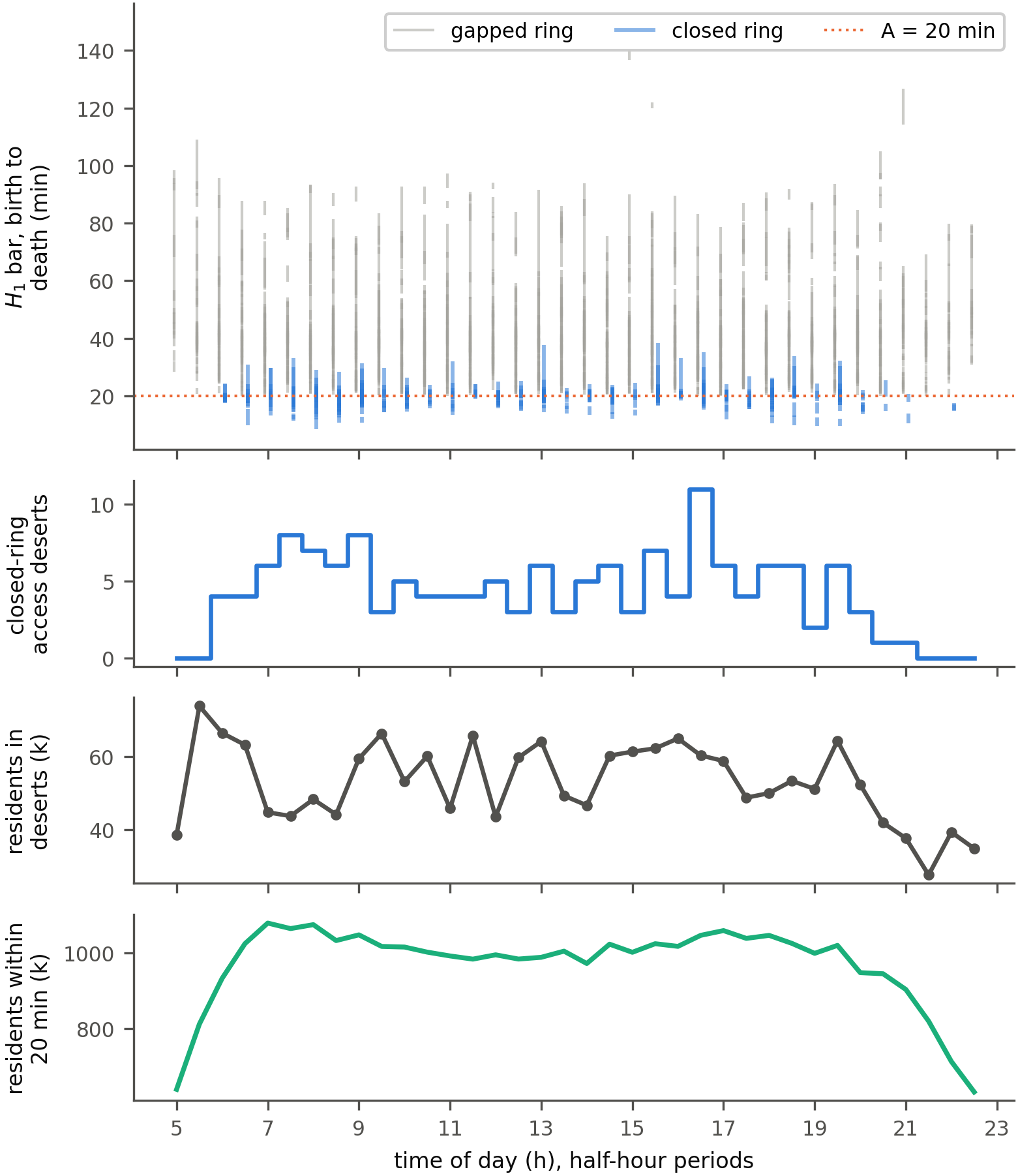}
\caption{Access-time persistence over the July 2020 weekdays in half-hour periods. Top: $H_1$ bars from birth to death (blue: closed ring; dotted line: $A = 20$ minutes). Below: closed-ring access deserts, residents in enclosed access deserts, and residents with an access time of at most 20 minutes.}
\label{fig:vineyard}
\end{figure}

Within the baseline weekdays (Figure~\ref{fig:vineyard}), the number of residents with an access time of at most 20 minutes lies between 971,870 and 1,078,242 in every half hour from 07:00 to 19:00, with the maximum at 07:00. After 20:00 this number falls, and it reaches 633,930 by 22:30. Between 07:00 and 19:00, the number of closed-ring access deserts varies between 3 and 11, and the number of residents in enclosed access deserts varies between 43,476 and 66,440. Each half-hourly diagram is based on 30 minutes of service per day. For this reason, we use the multi-hour periods of Figure~\ref{fig:windows} for comparisons, and we use the vineyard to show the daily pattern.

\subsection{Comparison with random reductions}\label{sec:res-null}

\begin{figure}[!t]
\centering
\includegraphics[width=\textwidth]{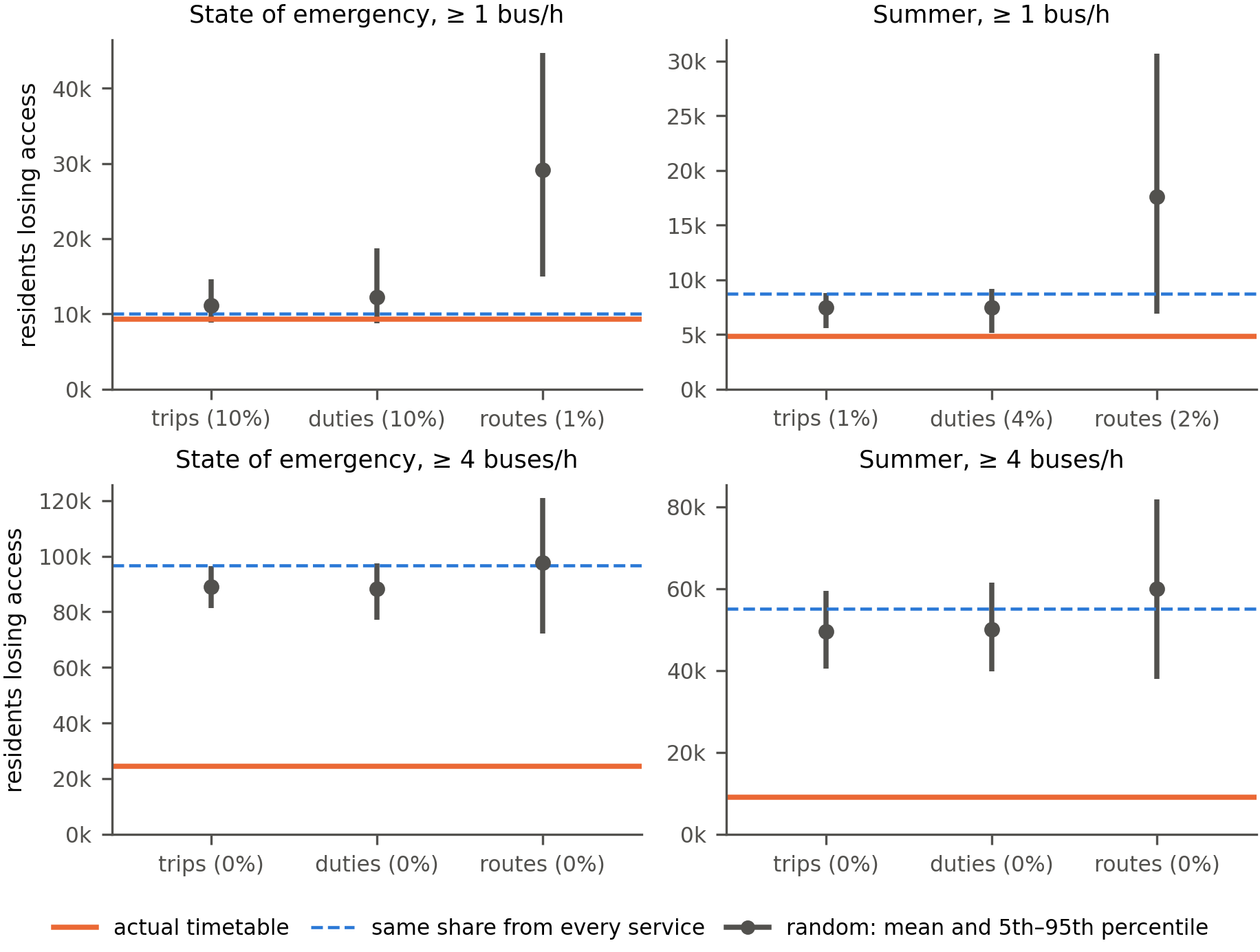}
\caption{Residents losing access at one and four buses an hour under each actual timetable (orange line), under 1,000 random reductions of each kind with the same moving bus-hours (mean and 5th--95th percentile), and when the same share of passes is removed from every service (dashed line). Tick labels give the share of random reductions that did as well as the actual timetable. The reductions are measured in moving bus-hours (22.6\% and 15.0\%); the same reductions are 17.7\% and 12.5\% of trip-unit bus-hours in Figure~\ref{fig:pareto}.}
\label{fig:null}
\end{figure}

Both reductions protected access better than chance (Table~\ref{tab:null}, Figure~\ref{fig:null}). At two and four buses an hour, every one of the 6,000 random reductions left more residents without service than the actual timetable did. Removing the same share of passes from every service would have left 96,701 residents without four buses an hour in the state of emergency and 55,086 in the summer, whereas the actual timetables left 24,544 and 9,041. At one bus an hour, the summer timetable did better than 2,930 of the 3,000 random reductions. The state-of-emergency timetable did better than 90\% of the random trip reductions, 90\% of the random duty reductions and 99\% of the random route reductions.

\begin{table}[!t]
\centering
\caption{Residents losing access under each actual timetable and under 1,000 random reductions of each kind with the same moving bus-hours, in the grid model. Each random-reduction cell gives the mean and, below it, the 5th--95th percentile; ``as good'' is the share of random reductions that left no more residents without service than the actual timetable. The shares of bus-hours in this table are shares of moving bus-hours (time in motion between GPS records); the same two reductions are 17.7\% and 12.5\% of trip-unit bus-hours (presence in 10-minute slots) in Table~\ref{tab:pareto}.}
\label{tab:null}
\small
\setlength{\tabcolsep}{4.5pt}
\begin{tabular}{@{}lrrcrcrc@{}}
\toprule
 & & \multicolumn{2}{c}{Random trips} & \multicolumn{2}{c}{Random duties} & \multicolumn{2}{c@{}}{Random routes} \\
\cmidrule(lr){3-4}\cmidrule(lr){5-6}\cmidrule(l){7-8}
Level & Actual & Mean (5--95\%) & \makecell{As\\good} & Mean (5--95\%) & \makecell{As\\good} & Mean (5--95\%) & \makecell{As\\good} \\
\midrule
\multicolumn{8}{@{}l}{\textit{State of emergency, May 2020: 22.6\% of moving bus-hours}}\\
$\geq 1$ bus/h & 9,307 & \makecell[r]{11,164\\(8,879--14,607)} & 0.10 & \makecell[r]{12,223\\(8,790--18,719)} & 0.10 & \makecell[r]{29,123\\(14,947--44,660)} & 0.01 \\
$\geq 2$ buses/h & 16,073 & \makecell[r]{39,191\\(33,341--45,573)} & 0.00 & \makecell[r]{39,615\\(33,001--46,309)} & 0.00 & \makecell[r]{46,967\\(30,635--64,343)} & 0.00 \\
$\geq 4$ buses/h & 24,544 & \makecell[r]{89,118\\(81,351--96,317)} & 0.00 & \makecell[r]{88,272\\(77,112--97,355)} & 0.00 & \makecell[r]{97,622\\(72,292--120,863)} & 0.00 \\
\multicolumn{8}{@{}l}{\textit{Summer school holiday, August 2020: 15.0\% of moving bus-hours}}\\
$\geq 1$ bus/h & 4,848 & \makecell[r]{7,484\\(5,563--8,806)} & 0.01 & \makecell[r]{7,460\\(5,145--9,161)} & 0.04 & \makecell[r]{17,633\\(6,895--30,693)} & 0.02 \\
$\geq 2$ buses/h & 9,877 & \makecell[r]{23,168\\(20,217--26,185)} & 0.00 & \makecell[r]{23,479\\(19,638--28,155)} & 0.00 & \makecell[r]{28,748\\(16,808--41,221)} & 0.00 \\
$\geq 4$ buses/h & 9,041 & \makecell[r]{49,630\\(40,522--59,417)} & 0.00 & \makecell[r]{50,127\\(39,849--61,479)} & 0.00 & \makecell[r]{59,942\\(38,027--81,740)} & 0.00 \\
\bottomrule
\end{tabular}
\end{table}

The diagrams show the same pattern. At one bus an hour, both actual timetables moved the desert diagram by a bottleneck distance of 194~m. In the state of emergency, every one of the 1,000 random trip, duty and route reductions of the same size moved it further. In the summer, the random trip reductions also all moved it further, and the random duty and route reductions matched or beat the actual timetable in only 1\% and 2\% of draws, respectively. On average, the random trip reductions moved the diagram by 545~m in the state of emergency and by 451~m in the summer. The reductions of the operator thus kept the overall shape of access closer to the baseline than chance would. However, they also did something that chance rarely does. Under both timetables, the number of residents inside closed-ring deserts at one bus an hour rose from 19,990 to 21,799 in this grid model, an increase of 9\% (Section~\ref{sec:res-topoaudit} compares these counts with those of the other models). Among the 2,000 random trip and duty reductions in each period, none raised this number as high as the actual timetables did. The highest 95th percentile among them, 20,491 under random duties in the state of emergency, was still only 2.5\% above the baseline. The fact that the increase is identical under the two timetables points to the same stops, and Section~\ref{sec:res-anatomy} shows that the two timetables thinned largely the same places.

\subsection{Trade-off frontier}\label{sec:res-pareto}

Random reductions are a low bar, and the frontier model sets a higher one. If trip units are removed in the order that costs the least access, a greedy search can remove 27.1\% of morning bus-hours before the best stop of any resident falls below one, two or four buses an hour. If every route must also stay within capacity, this share is 19.8--20.5\%. These shares take only the thresholds into account, while the frontier also takes into account the waiting time of riders.

\begin{figure}[!t]
\centering
\includegraphics[width=\textwidth]{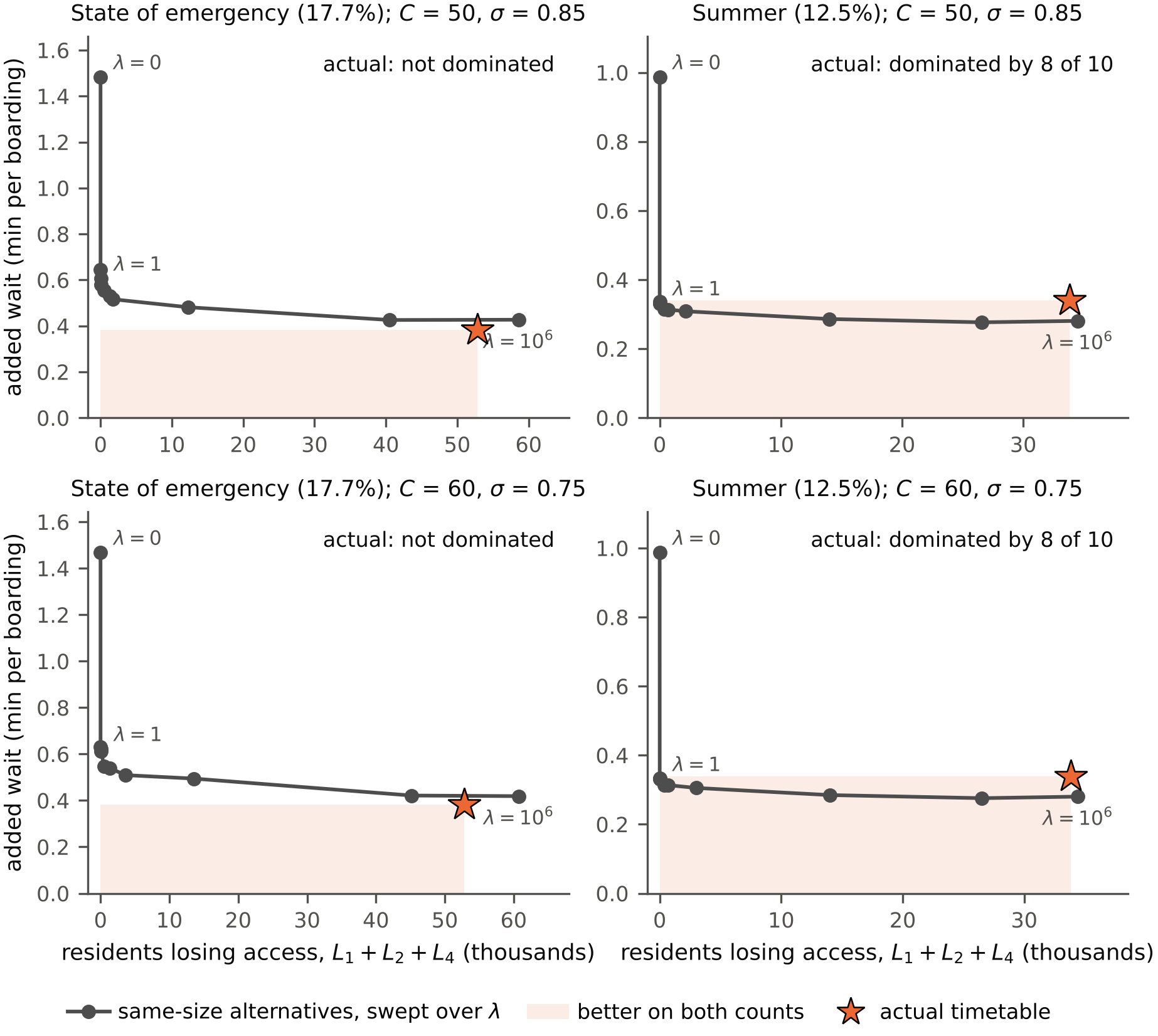}
\caption{Residents losing access (summed over one, two and four buses an hour) against added waiting per boarding for capacity-feasible alternatives of the same size, swept over $\lambda$ (dots), and for the actual timetable (star). The shaded region contains the alternatives that are no worse on either count. Rows: vehicle capacity of 50 boardings per run with a card share of 0.85, and 60 with 0.75. $L_F$ is the number of residents who have service at $F$ or more buses an hour within 800~m in the baseline but no longer do. The sum $L_1 + L_2 + L_4$ counts threshold crossings, not unique residents, because a resident who loses all service is counted at all three levels. Sizes are shares of trip-unit bus-hours (17.7\% and 12.5\%), which correspond to 22.6\% and 15.0\% of moving bus-hours in Figure~\ref{fig:null}.}
\label{fig:pareto}
\end{figure}

At the size of each actual reduction, there is a capacity-feasible alternative that leaves every resident above all three thresholds. Taken alone, however, this access-first alternative removes service from the busiest corridors, where most riders wait. It adds 1.47--1.48 minutes of waiting per boarding at the size of the state-of-emergency reduction and 0.99 minutes at the size of the summer reduction, compared with 0.38 and 0.34 minutes for the actual timetables. Sweeping $\lambda$ reveals the alternatives between these two extremes (Figure~\ref{fig:pareto}, Table~\ref{tab:pareto}).

\begin{table}[!t]
\centering
\caption{Actual timetables and capacity-feasible alternatives of the same size in the frontier model: residents losing access at one, two and four buses an hour, and added expected waiting per boarding and per resident. Ranges span the two settings of vehicle capacity and card share; single values are identical in both. $L_F$ is the number of residents who have service at $F$ or more buses an hour within 800~m in the baseline but no longer do after the reduction. The percentages are shares of trip-unit bus-hours (presence in 10-minute slots); the same two reductions are 22.6\% and 15.0\% of moving bus-hours (time in motion) in Table~\ref{tab:null}.}
\label{tab:pareto}
\small
\begin{tabular}{@{}lrrrcc@{}}
\toprule
 & & & & \multicolumn{2}{c@{}}{Added wait (min)} \\
\cmidrule(l){5-6}
Plan & $L_1$ & $L_2$ & $L_4$ & per boarding & per resident \\
\midrule
\multicolumn{6}{@{}l}{\textit{State of emergency (17.7\%)}}\\
actual & 10,006 & 18,230 & 24,621 & 0.38 & 1.76 \\
access first ($\lambda = 0$) & 0 & 0 & 0 & 1.47--1.48 & 0.99--1.18 \\
$\lambda = 1$ & 0 & 4 & 40--62 & 0.61--0.62 & 1.02--1.03 \\
$\lambda = 300$ & 3,355--3,730 & 4,269 & 4,646--5,530 & 0.48--0.49 & 1.28--1.30 \\
waiting first ($\lambda = 10^6$) & 16,985 & 17,246--19,382 & 24,376--24,387 & 0.42--0.43 & 1.91 \\
\multicolumn{6}{@{}l}{\textbf{Not dominated}: every alternative with less access loss has more added waiting} \\
\midrule
\multicolumn{6}{@{}l}{\textit{Summer (12.5\%)}}\\
actual & 5,253 & 18,428 & 10,150 & 0.34 & 1.26 \\
access first ($\lambda = 0$) & 0 & 0 & 0 & 0.99 & 0.73 \\
$\lambda = 0.3$--1 & 0 & 0 & 0--2 & 0.33--0.34 & 0.58--0.59 \\
$\lambda = 10$ & 0 & 62 & 325 & 0.31 & 0.59 \\
$\lambda = 100$ & 674--1,493 & 428--578 & 964--1,034 & 0.31 & 0.67 \\
$\lambda = 300$ & 1,580 & 4,138 & 8,256--8,270 & 0.28--0.29 & 0.74--0.75 \\
\multicolumn{6}{@{}l}{\textbf{Dominated} by 8 of the 10 alternatives in both settings} \\
\bottomrule
\end{tabular}
\end{table}

\paragraph{State-of-emergency timetable.} None of the ten alternatives had both less access loss and less added waiting. The alternatives that kept hourly service for nearly every resident ($\lambda = 1$) added 0.61--0.62 minutes per boarding. The alternatives that minimized waiting still added 0.42--0.43 minutes, and they left more residents without hourly service than the actual timetable did. The operator chose short waits for the passengers who were still traveling. The cost of the other choice can be read from the swept alternatives: keeping hourly service for all 10,006 residents would have added 0.23--0.24 minutes to each of the 1,646--1,865 morning boardings, which is about 380 to 450 minutes of waiting per weekday across all riders.

\paragraph{Summer timetable.} Eight of the ten alternatives dominate it in both settings. For $\lambda = 0.3$ to 3, the alternatives left no resident without hourly or half-hourly service, and at most 37 residents below four buses an hour, with the same added waiting per boarding as the actual timetable (0.33--0.34 minutes). For $\lambda = 10$ to 300, the alternatives reduced the added waiting to 0.28--0.31 minutes per boarding, which is a reduction of 9--18\% in the added expected waiting that the reduction causes (from 0.34 minutes under the actual timetable), not a reduction in total expected waiting, and they still left fewer residents without service than the actual timetable at every level. The added waiting per resident weights each stop by the people who live near it rather than by those who ride. This quantity falls from 1.26 minutes under the actual timetable to 0.58--0.75 minutes under these alternatives.

Both verdicts are the same under the two settings of capacity and card share. The access figures for the actual timetables and for the access-first alternative are identical in the two settings, and those of the intermediate and waiting-first alternatives differ by the ranges shown in Table~\ref{tab:pareto}, for example $L_1 = 3{,}355$--$3{,}730$ at $\lambda = 300$ under the state-of-emergency reduction. The topology of the alternatives confirms the difference between the two timetables. At four buses an hour, the summer timetable moved the diagram of enclosed deserts by a bottleneck distance of 437~m ($W_1$ 2.2~km), whereas an access-first greedy reduction of the same size, computed without the capacity constraint, left it unchanged. At one bus an hour, the actual timetable moved the diagram by 201~m and reduced the number of enclosed deserts from 35 to 34, while the greedy reduction left all 35 in place. Under a reduction, a smaller count means that a ring has opened or that two pockets have merged. The state-of-emergency timetable moved the four-bus diagram by the same 437~m and reduced its number of deserts from 18 to 16.

\subsection{Spatial pattern of the reductions}\label{sec:res-anatomy}

\begin{figure}[!t]
\centering
\includegraphics[width=0.7\textwidth]{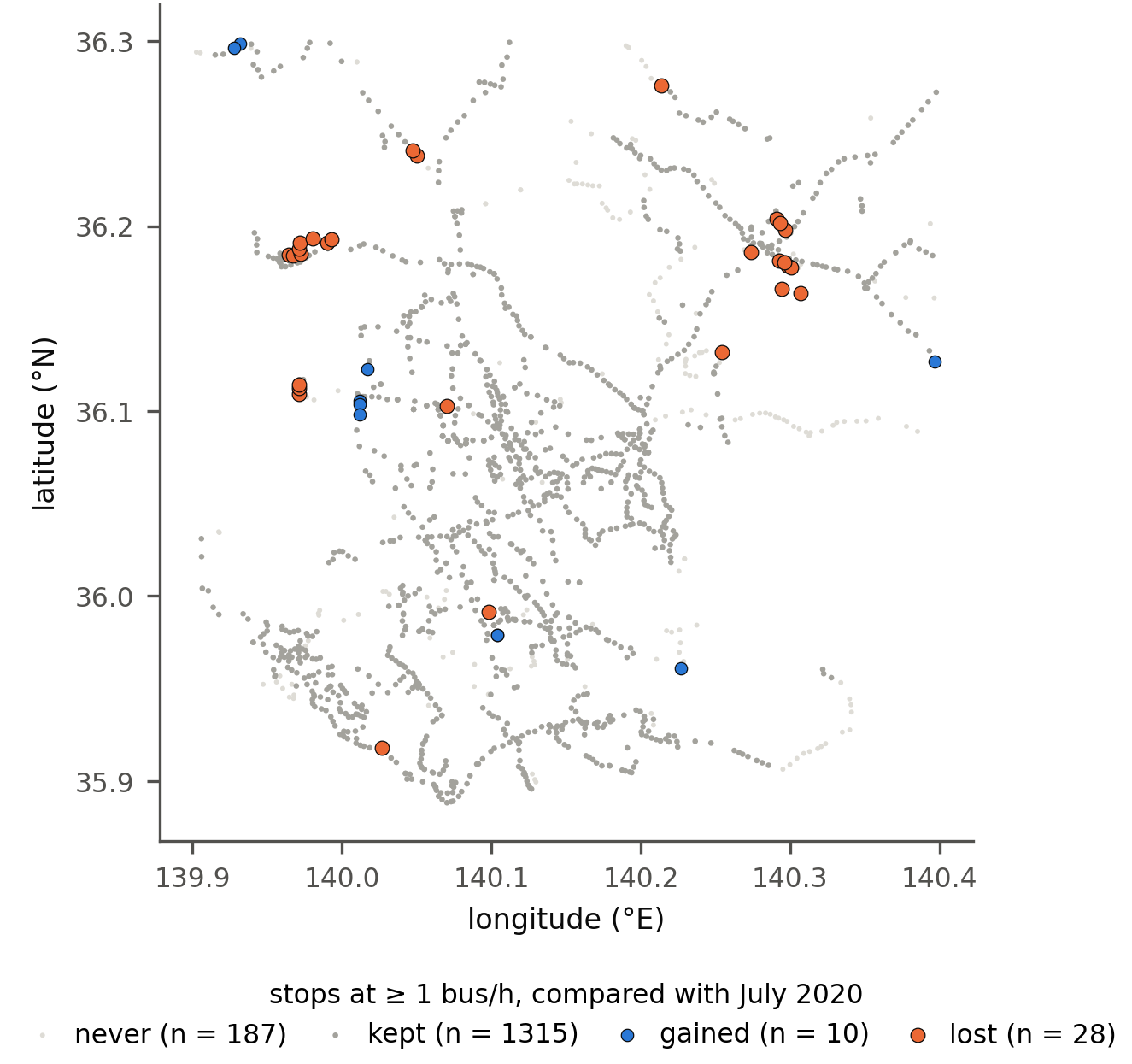}
\caption{Stops that keep, lose or gain at least one bus an hour in the weekday morning peak under the August 2020 summer timetable, relative to July 2020.}
\label{fig:cutmap}
\end{figure}

The stop-level record explains the two verdicts. Under the summer timetable, 28 of the 1,540 stops fell below one bus an hour, 10 rose above it and 1,315 stayed above it (Figure~\ref{fig:cutmap}). The ten that rose all had a July rate of 0.81--0.98 an hour, just below the threshold, and none of them picked up a route that was not already at or near full retention between July and August, so their rise reflects ordinary variation from week to week rather than added service. The 28 stops were served 1.11 times an hour in July against 4.50 for the stops that kept hourly service (medians), and they registered a median of 0.00 card taps per weekday morning against 0.81. They were not, however, stops with few residents: a median of 3,377 people live within 800~m of them, against 3,340 for the stops that kept service. Even among stops in the same July frequency band of one to two buses an hour, the stops that lost hourly service had fewer riders (0.19 against 0.86 card taps per day) and more residents (3,377 against 1,529). The summer timetable thus withdrew hourly service where few people rode and as many people lived as elsewhere, which is the choice that a ridership criterion favors and a coverage criterion penalizes. Most of the 28 stops lie on outlying branches at the northern, northeastern and western edges of the network, a median of 17.8~km from Tsukuba Station against 11.7~km for the stops that kept service.

The state-of-emergency timetable withdrew hourly service from 41 stops. These too had little service and few riders (1.31 passes per hour and 0.05 morning card taps, medians), but, unlike the summer stops, they also had few residents: a median of 1,154 people within 800~m, against 3,398 around the stops that kept hourly service, at a median of 19.8~km from Tsukuba Station. This is consistent with the sweep finding no dominating alternative for it: it removed service where both riders and residents were scarce, so an alternative that spared these residents had to take the same bus-hours from stops where more people ride and would wait longer.

Beyond these threshold crossings, both timetables thinned service almost uniformly. The median stop kept 96\% of its morning service in the state of emergency and 95\% in the summer, with medians between 94\% and 98\% in every quartile of July frequency, and a route's July ridership did not predict how much of its service it kept (Spearman rank correlations of $-0.08$ and $-0.03$ over 258 route pairs). The two timetables also resemble each other. Across the 1,343 stops served at least hourly in July, the share of service each stop kept in the state of emergency correlates with the share it kept in the summer (Spearman rank correlation 0.63), and the six routes the summer timetable reduced most kept a median of 58\% of their service in the state of emergency against 97\% for other routes. Routes named after a school or university were not reduced more than others in the state of emergency (median retention 0.98 against 0.97).

\subsection{Distribution of added waiting}\label{sec:res-equity}

The averages of Section~\ref{sec:res-pareto} describe the typical rider. This section asks who bears the tail of the added wait. We measure the increase in expected wait $\Delta w_s$ at each stop that is served in July (Section~\ref{sec:equity}) and weight it by stops, by boardings in the window of the reduction and by residents.

\begin{figure}[!t]
\centering
\includegraphics[width=\textwidth]{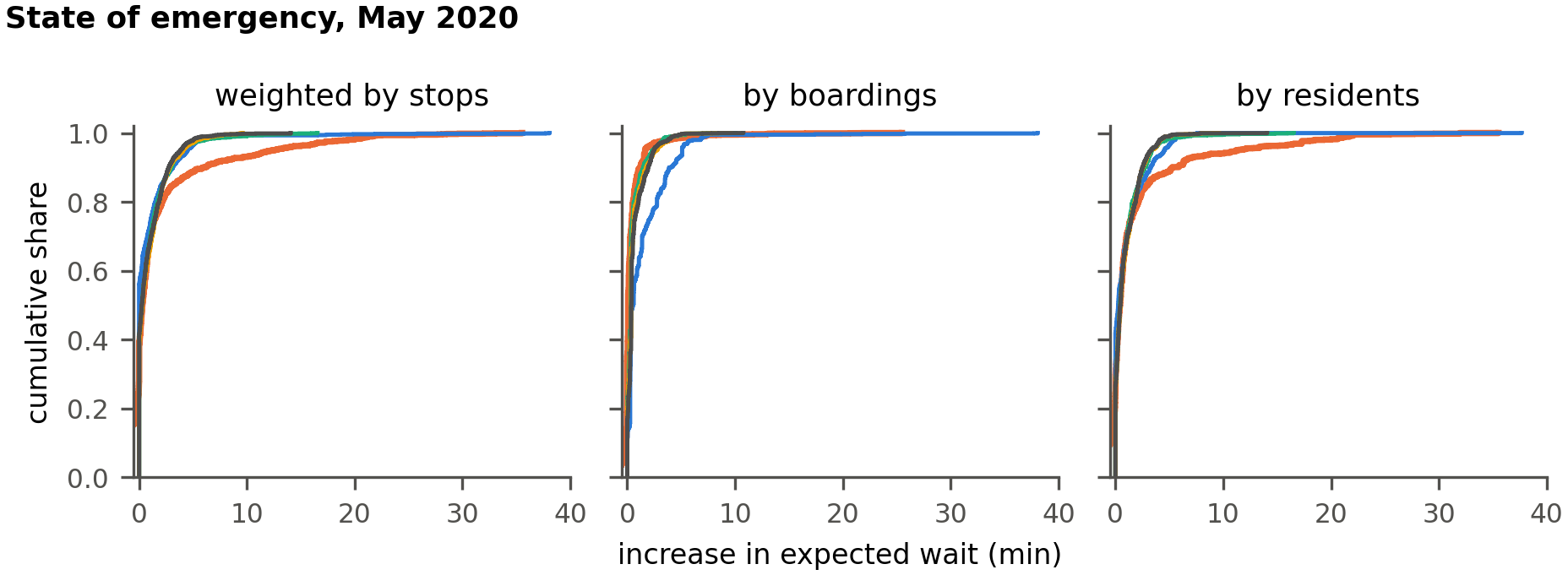}\par\vspace{4pt}
\includegraphics[width=\textwidth]{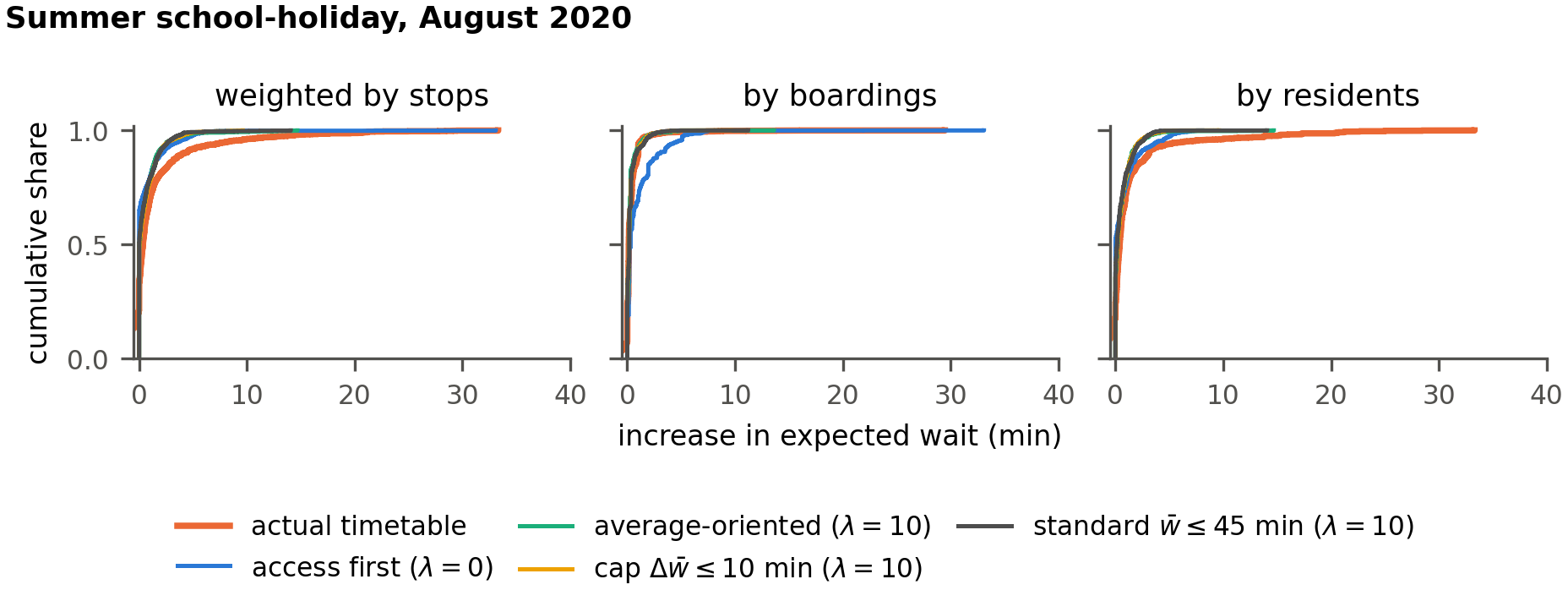}
\caption{Cumulative distribution of the increase in expected wait under the actual timetable (orange), the access-first plan ($\lambda = 0$, blue), the average-oriented plan ($\lambda = 10$, green) and two equity-constrained plans ($\Delta \bar w \leq 10$, amber, and $\bar w \leq 45$, dark gray, both with $\lambda = 10$), at the size of the state-of-emergency reduction (top; 17.7\% of trip-unit bus-hours) and of the summer reduction (bottom; 12.5\%), weighted by stops, boardings and residents. These are distributions of expected waits under the model, not of realized waits.}
\label{fig:equity-cdf}
\end{figure}

\begin{figure}[!t]
\centering
\includegraphics[width=\textwidth]{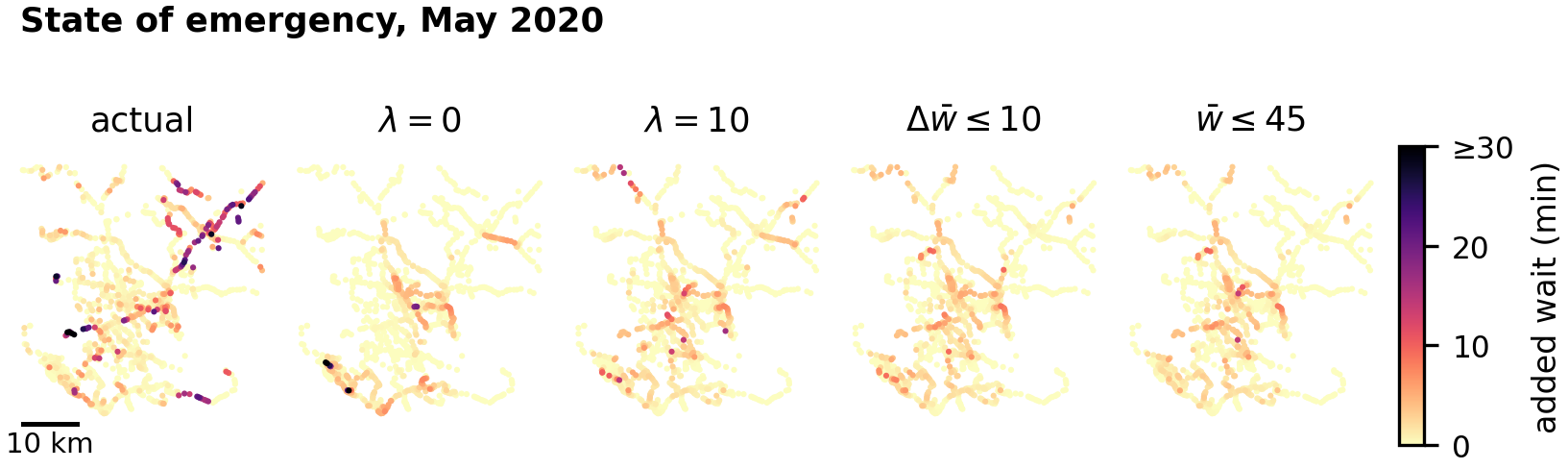}\par\vspace{4pt}
\includegraphics[width=\textwidth]{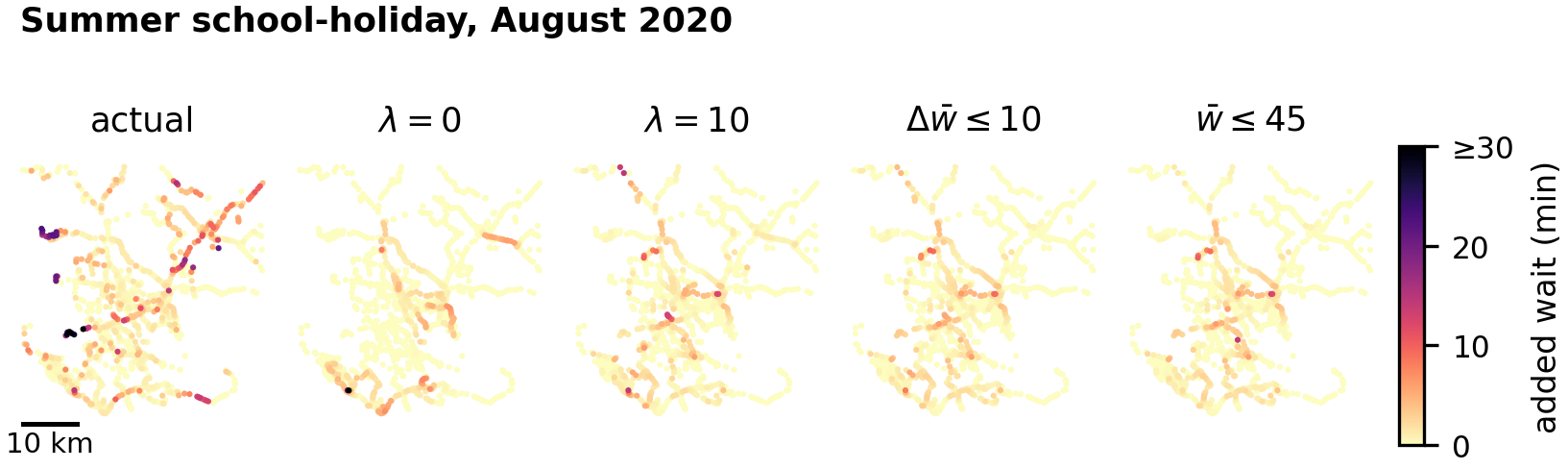}
\caption{Increase in expected wait at each stop (minutes, truncated at 30) under the actual timetable and four alternatives, at the size of the state-of-emergency reduction (top; 17.7\% of trip-unit bus-hours) and of the summer reduction (bottom; 12.5\%). From left to right: actual, access first ($\lambda = 0$), average-oriented ($\lambda = 10$), $\Delta \bar w \leq 10$ and $\bar w \leq 45$, both with $\lambda = 10$. The bar in the first panel is 10~km.}
\label{fig:equity-map}
\end{figure}

Table~\ref{tab:equity} gives the added expected wait of the actual timetables and of the alternatives of the same size, weighted by stops, boardings and residents, together with the upper tail. The mean per boarding under the actual timetables, 0.38 minutes in the state of emergency and 0.34 in the summer, hides a concentration. Weighted by residents, the same increases average 1.76 and 1.26 minutes, because the riders of those mornings were concentrated at stops that kept their service, whereas the stops that lost service had few riders. The 95th percentile of the added wait of residents is 11.4 and 6.4 minutes, the largest increase at any stop is 35.5 and 33.3 minutes, and 5.9\% and 3.9\% of residents are assigned to stops whose expected wait rises by 10 minutes or more (6.9\% and 3.8\% of stops, and 0.47\% and 0.19\% of boardings). For the alternatives with $\lambda = 10$, the resident-weighted mean is 0.97 and 0.59 minutes, the 95th percentile is 3.4 and 2.4 minutes, the maximum is 16.6 and 14.7 minutes, and 0.5\% and 0.3\% of residents are at stops with an increase of 10 minutes or more. The actual timetables therefore concentrate the burden more than the swept plans do, although in the state of emergency they have the lowest mean per boarding (0.38 against 0.55 minutes). Figure~\ref{fig:equity-cdf} shows the cumulative distributions and Figure~\ref{fig:equity-map} the location of the increases. Under the actual timetables the largest increases lie at stops on a few outlying corridors at the edge of the study area and at isolated stops in the west and southeast, whereas the increases of the swept plans are smaller and more widely spread.

The stops with an increase of 10 minutes or more (104 of the 1,507 stops served in the baseline in the state of emergency and 58 in the summer) lie far from the center, at a median of 19.5 and 17.2~km from Tsukuba Station against 12.2 and 12.5~km for the other served stops. They had less service before the reduction (2.0 and 1.6 passes per hour against 4.1 and 4.0), almost no morning card taps (0.10 and 0.00 against 0.57 and 0.57) and fewer residents within walking distance (1,131 and 2,029 against 3,184 and 3,052), and they kept a median of 60\% and 65\% of their service against 97\% and 95\%.

\begin{table}[!t]
\centering
\caption{Distribution of the increase in expected wait $\Delta w_s$ (minutes) under the actual timetable, the average-oriented alternatives ($\lambda = 10$) and the equity-constrained alternatives at the size of the state-of-emergency and summer reductions. $\Delta \bar w \leq 10$ and $\Delta \bar w \leq 20$ cap the increase at every protected stop; $\bar w \leq 45$ caps the wait after the reduction; all three use $\lambda = 10$. Means are weighted by stops, boardings and residents; the 95th percentile and the share with an increase of at least 10 minutes are weighted by residents. ``Max removable'' is the largest share of trip-unit bus-hours that can be removed while meeting the constraint (sweep limit 40\%; values near 40\% mean that the constraint did not bind). $L_1 + L_2 + L_4$ counts threshold crossings (Section~\ref{sec:alternatives}). Protected set: all stops served in the baseline.}
\label{tab:equity}
\small
\setlength{\tabcolsep}{4.5pt}
\begin{tabular}{@{}lrrrrrrrr@{}}
\toprule
 & & & \multicolumn{3}{c}{Mean of $\Delta w_s$} & \multicolumn{2}{c}{Residents} & \\
\cmidrule(lr){4-6}\cmidrule(lr){7-8}
Plan & \makecell[r]{Max\\removable} & \makecell[r]{$L_1 + L_2$\\${}+ L_4$} & stops & \makecell[r]{board-\\ings} & \makecell[r]{resi-\\dents} & \makecell[r]{95th\\pct} & \makecell[r]{$\geq 10$\\min} & \makecell[r]{Max\\(stops)} \\
\midrule
\multicolumn{9}{@{}l}{\textit{State of emergency (17.7\%)}}\\
actual & -- & 52,856 & 1.68 & 0.38 & 1.76 & 11.4 & 5.93\% & 35.5 \\
$\lambda = 10$ & -- & 534 & 1.00 & 0.55 & 0.97 & 3.4 & 0.52\% & 16.6 \\
$\Delta \bar w \leq 10$ & 29.5\% & 521 & 0.99 & 0.64 & 0.93 & 3.3 & 0.00\% & 9.6 \\
$\Delta \bar w \leq 20$ & 37.5\% & 566 & 0.97 & 0.58 & 0.93 & 3.3 & 0.39\% & 18.7 \\
$\bar w \leq 45$ & 37.0\% & 547 & 0.94 & 0.68 & 0.92 & 3.3 & 0.05\% & 14.1 \\
\multicolumn{9}{@{}l}{\textit{Summer (12.5\%)}}\\
actual & -- & 33,831 & 1.27 & 0.34 & 1.26 & 6.4 & 3.90\% & 33.3 \\
$\lambda = 10$ & -- & 388 & 0.64 & 0.31 & 0.59 & 2.4 & 0.27\% & 14.7 \\
$\Delta \bar w \leq 10$ & 27.8\% & 371 & 0.62 & 0.33 & 0.55 & 2.2 & 0.00\% & 9.6 \\
$\Delta \bar w \leq 20$ & 35.6\% & 388 & 0.65 & 0.32 & 0.58 & 2.5 & 0.39\% & 18.7 \\
$\bar w \leq 45$ & 33.2\% & 371 & 0.60 & 0.34 & 0.54 & 2.4 & 0.04\% & 14.1 \\
\bottomrule
\end{tabular}
\end{table}

\begin{figure}[!t]
\centering
\includegraphics[width=\textwidth]{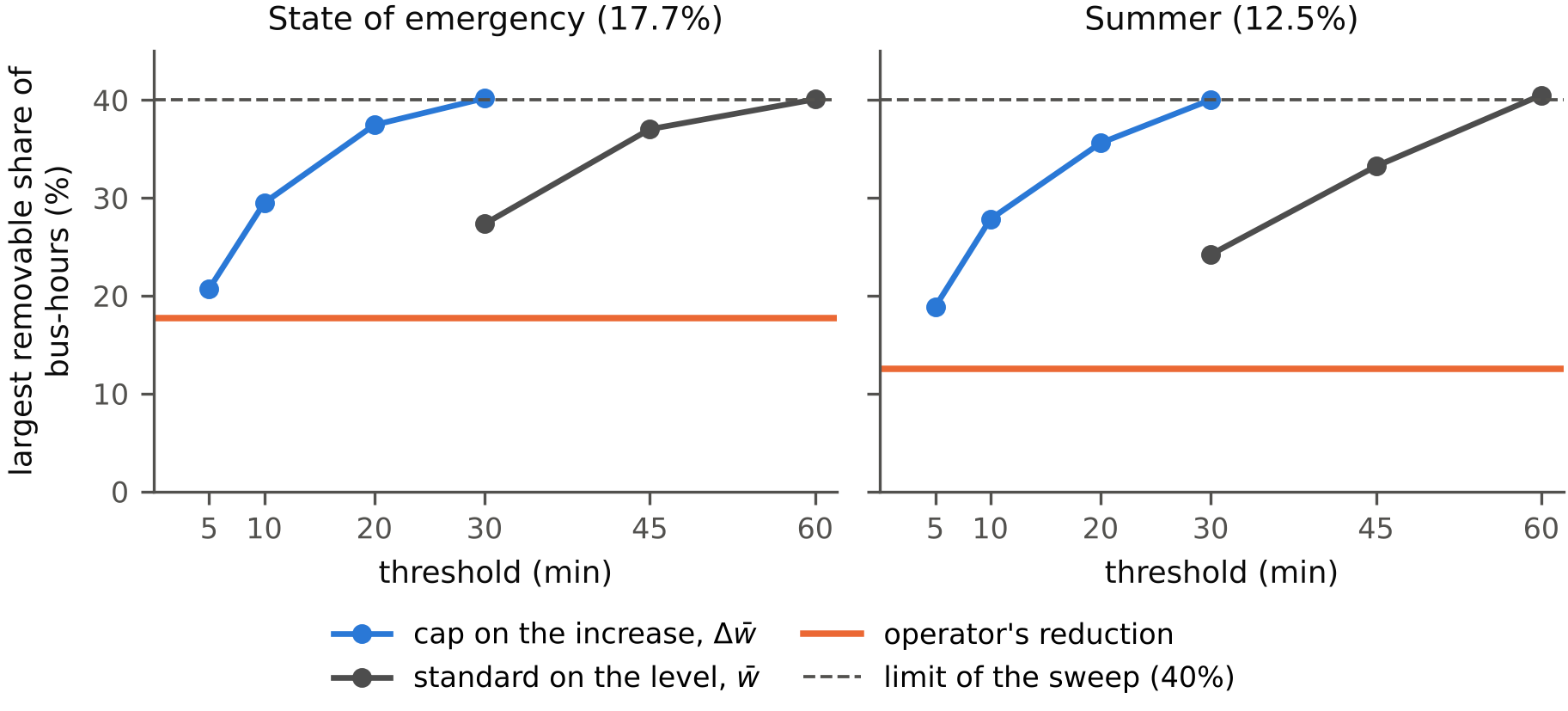}
\caption{Largest share of trip-unit bus-hours that can be removed while every stop served in the baseline meets the standard, at the size of the state-of-emergency reduction (left) and of the summer reduction (right), with $\lambda = 10$. Blue: cap on the increase in expected wait ($\Delta \bar w$); dark gray: standard on the wait after the reduction ($\bar w$). Orange line: the size of the operator's reduction. Dashed line: the limit of the sweep (40\%); a point on that line means that the standard did not bind.}
\label{fig:standards}
\end{figure}

\paragraph{Cost of the standards.} Every standard tested can be met at the size of the operator’s reduction (Figure~\ref{fig:standards}). When only the 14 stops with at least 20 morning boardings in the state-of-emergency window are protected, no standard binds within the swept range (the largest removable share is 40.0\% to 40.4\% for every standard), because these busy stops keep frequent service in every plan. The results below therefore protect all stops served in the baseline. With $\lambda = 10$, the largest share of bus-hours that can be removed while keeping the increase at every protected stop within 5 minutes is 20.7\% in the state-of-emergency data and 18.9\% in the summer data, against budgets of 17.7\% and 12.5\%. For $\Delta \bar w \leq 10$ the shares are 29.5\% and 27.8\%, and for $\bar w \leq 45$ they are 37.0\% and 33.2\%. A value near 40\% means that the standard did not bind within the swept range.

Tightening the standard to 10 minutes costs little at this size. In the state of emergency the $\Delta \bar w \leq 10$ plan has a maximum increase of 9.6 minutes instead of 16.6, a mean of 0.64 minutes per boarding instead of 0.55, and 521 instead of 534 residents in the $L_1 + L_2 + L_4$ count. The $\Delta \bar w \leq 5$ plan costs more (0.75 minutes per boarding and 727 residents). In the summer the $\Delta \bar w \leq 10$ plan raises the mean from 0.31 to 0.33 minutes per boarding and leaves fewer residents below the thresholds (371 against 388). Most of the plan is unchanged: its removed bus-hours coincide with those of the unconstrained plan at a bus-hour Jaccard index of 78\% (state of emergency) and 81\% (summer), and at 87\% and 90\% for $\Delta \bar w \leq 20$. The access-first plan ($\lambda = 0$) shares only 36\% and 19\% of its bus-hours with the average-oriented plan. In the average-oriented plan, 68\% (state of emergency) and 70\% (summer) of the removed bus-hours lie between 07:30 and 08:30. Imposing the $\Delta \bar w \leq 10$ standard keeps 9.5 of the 81.5 removed bus-hours in service in the state of emergency (21 trip units on 18 route pairs) and removes other trips in their place, and the corresponding figures for the summer are 6.2 of 57.8 bus-hours (13 units on 12 route pairs). The 07:30--08:00 block carries 39\% of the removed bus-hours without the standard and 40\% with it in the state of emergency, and 34\% and 36\% in the summer, so the standard does not move the reduction systematically toward the edges or the center of the peak.

The route pairs that keep most of the service they would have lost are 96054 (unnamed), 71910 (a service to and from Tsukuba Center) and 60206 (a direct school service from Tsuchiura Station East Exit to Tsuchiura Nihon University High School) in the state of emergency, and 35001 (unnamed), 96054 (unnamed) and 68001 (a loop from Arakawaoki Station East Exit to the Ami Premium Outlets) in the summer. The trips removed in their place come mainly from 60802 (Arakawaoki Station to Tsukuba Center via Gakuen-Namiki), 60801 (Arakawaoki Station to the University of Tsukuba via Gakuen-Namiki and Tsukuba Center) and 60708 (Tsukuba Center to Tsukuba Techno Park Oho) in the state of emergency, and from 70202 (unnamed), 60708 and 95006 (a service to and from Tsuchiura Station) in the summer. Route pairs are identified by their codes in the smart-card records, with names from the operator's route-number table (2020) and the inferred route endpoints; pairs marked unnamed are listed in neither.

\subsection{Financial balance}\label{sec:res-finance}

Revenue per trip unit is taken from the IC taps of the baseline window, divided by the card share ($\sigma = 0.75$), and multiplied by the ratio of boardings in the window of the reduction to those in the baseline, 0.344 in the state of emergency and 0.784 in the summer; the mean is 276 yen per boarding. The tap fare field is ambiguous between a per-record and a per-passenger amount (102 of 131,923 records carry more than one passenger, and the median fare of two-passenger records is 1.5 times that of one-passenger records on the same route and stop), which affects about 0.2\% of revenue.

\begin{figure}[!t]
\centering
\includegraphics[width=\textwidth]{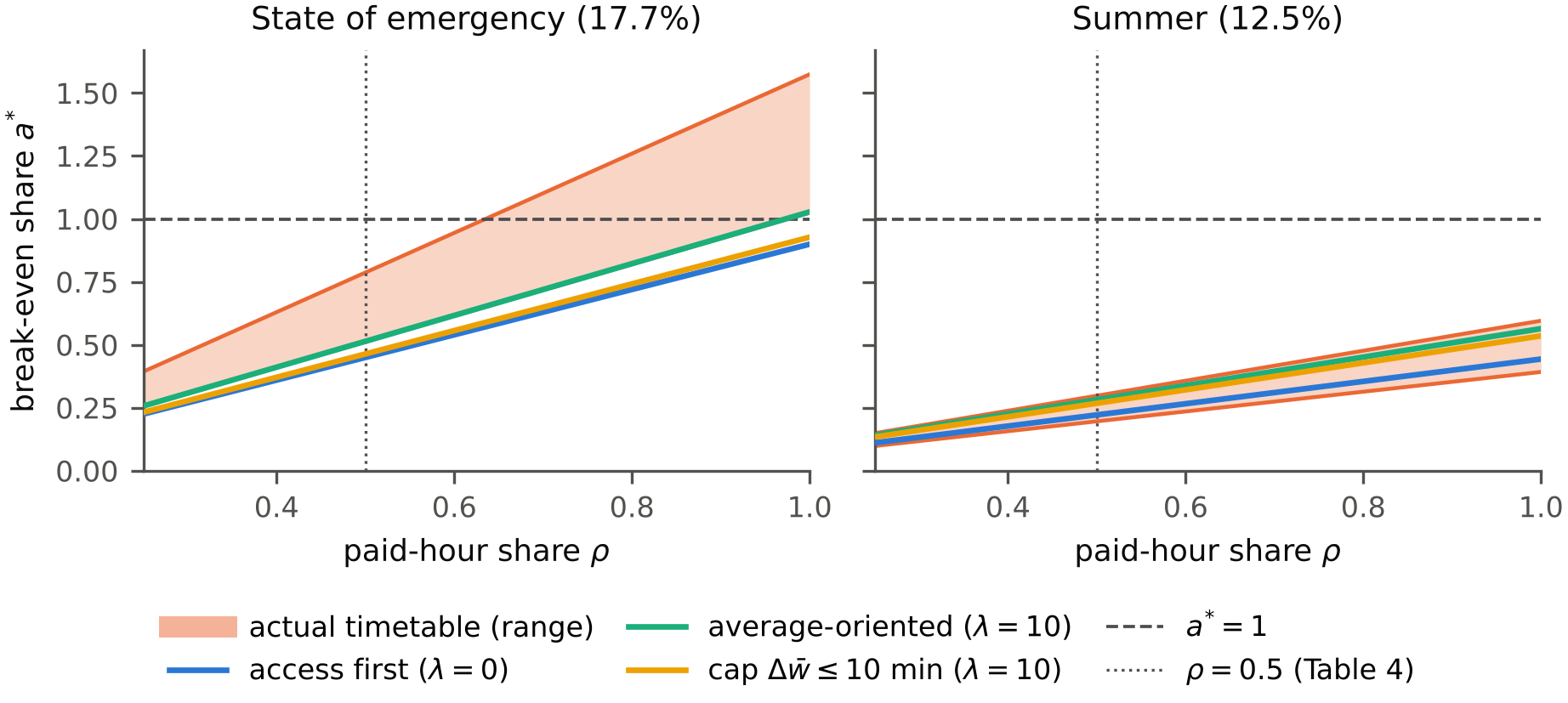}
\caption{Break-even abandonment share $a^*$ against the share $\rho$ of removed bus-hours that become paid hours saved, at $c_h = $ JPY 1,906 per hour, for the plans of Table~\ref{tab:finance}, in the state of emergency (left) and the summer (right); the shaded band is the range of the actual timetable. Above the dashed line ($a^* = 1$) profit improves even if every passenger on the removed trips stops paying; the dotted line marks $\rho = 0.5$, the value used in Table~\ref{tab:finance}. The reductions are 17.7\% and 12.5\% of trip-unit bus-hours. $a^*$ is proportional to $\rho$ and to $c_h$.}
\label{fig:breakeven}
\end{figure}

Table~\ref{tab:finance} reports, for the operator's reductions and for the alternatives, the operated hours removed, the revenue on the removed trip units and the break-even abandonment share $a^*$ at the benchmark wage, and Figure~\ref{fig:breakeven} shows how $a^*$ depends on $\rho$. The operator's reductions are given as a range. Their trip units with a route label lost 125.1 bus-hours in the state of emergency and 87.7 in the summer, more than the 81.4 and 57.7 bus-hours that the timetables saved, because trips also moved between 30-minute blocks and the hours of service without a route label rose. The upper value of the range is the revenue on all labeled bus-hours lost (the ``gross'' rows of Table~\ref{tab:finance}), and the lower value scales it to the bus-hours saved (the ``net'' rows). In the state of emergency the actual timetable put JPY 98,614 to 151,502 of revenue at risk each weekday morning, less than the alternatives of the same size at the lower value and about as much as the average-oriented plan at the upper value (151,147 for the average-oriented plan, 169,396 for the equity-constrained plan and 172,757 for the access-first plan). Its $a^*$ at $\rho = 0.5$ is 0.51 to 0.79, against 0.51, 0.46 and 0.45. In the summer the range is JPY 184,781 to 280,634, against 195,190, 205,831 and 250,084, so the actual timetable put about as much revenue at risk as the average-oriented and equity-constrained plans at the lower value and the most of the four plans at the upper value; its $a^*$ at $\rho = 0.5$ is 0.20 to 0.30, against 0.28, 0.27 and 0.22. At $\rho = 0.5$ no plan reaches $a^* = 1$ in either window: if half of the removed hours become paid hours saved, the labor saving covers at most about four fifths of the revenue on the removed trips, so each plan improves profit only if enough passengers keep paying. At $\rho = 1$ the state-of-emergency timetable (1.02 to 1.57) and the average-oriented plan (1.03) reach 1, and no summer plan does (at most 0.60). The break-even scales with the wage: for the actual timetable at $\rho = 0.5$ it is 0.67 to 1.03 and 0.86 to 1.32 (state of emergency) and 0.26 to 0.39 and 0.33 to 0.50 (summer) at hourly costs of JPY 2,500 and 3,200.

Random removals of the same size carry JPY 1,605 per removed bus-hour in the state of emergency (5th to 95th percentile 1,404 to 1,813) and 3,681 in the summer (3,111 to 4,341). The actual timetables removed 1,211 to 1,860 and 3,200 to 4,860, so depending on the attribution they removed less or more revenue per bus-hour than random removals, and the data do not show that the operator selected trips with little revenue. The access-first plan removed 2,119 and 4,299, and the average-oriented plan 1,854 and 3,378. At $\rho = 0.5$ and retention $\theta = 0.8$ the change in profit is positive for all four plans in the state of emergency (JPY 47,318 to 57,896 for the actual timetable, and 43,148, 47,448 and 44,675 for the alternatives in the order of Table~\ref{tab:finance}). In the summer it is $-$1,092 to 18,079 for the actual timetable and 5,424, 16,024 and 13,941 for the alternatives, so the equity-constrained plan, which meets the 10-minute standard, improves profit in both windows. At $\rho = 0.25$ the change is positive for every plan in the state of emergency (JPY 8,509 to 19,087 for the actual timetable) and negative for every plan in the summer. The $(1 - \rho) \Delta H$ operated hours that do not become paid-hour savings, 40.7 hours (state of emergency) and 28.9 hours (summer) per weekday morning at $\rho = 0.5$, are released capacity and are not counted as savings.

The revenue-aware plans show the tension with access. With the heaviest revenue weight (0.1 residents per yen), $R$ falls to JPY 63,870 and 59,964 and $a^*$ at $\rho = 0.5$ rises to 1.24 and 0.92, but the added wait per boarding rises to 1.27 and 0.66 minutes, and 606 and 658 residents fall below the thresholds. Because the labor saving is the same for plans of equal size, the plans differ only in the revenue they put at risk. These figures cover the weekday morning peak only and are not annualized.

\begin{table}[!t]
\centering
\caption{Financial comparison per weekday morning peak (07:00--09:00). $\Delta H$: operated bus-hours removed. $R$: fare revenue on the removed trip units, from the baseline taps scaled to the demand of the window of the reduction. The actual timetables have two rows (Section~\ref{sec:res-finance}): ``net'' scales the revenue to the bus-hours saved, and ``gross'' counts all labeled bus-hours lost. $a^*$: break-even abandonment share, $c_h \rho \, \Delta H / R$, at $c_h = $ JPY 1,906 per hour and the stated $\rho$; $a^* \geq 1$ means that profit improves even if every passenger on the removed trips stops paying. $\Delta \Pi$: change in profit per weekday morning at $\rho = 0.5$ and retention $\theta = 0.8$. Fares are held fixed. The equity-constrained plan is the $\Delta \bar w \leq 10$ plan with $\lambda = 10$. The percentages are shares of trip-unit bus-hours (22.6\% and 15.0\% of moving bus-hours in Table~\ref{tab:null}).}
\label{tab:finance}
\small
\begin{tabular}{@{}lrrrrrr@{}}
\toprule
 & & & \multicolumn{3}{c}{$a^*$} & \\
\cmidrule(lr){4-6}
Plan & $\Delta H$ (h) & $R$ (JPY) & $\rho = 0.25$ & $\rho = 0.5$ & $\rho = 1$ & $\Delta \Pi$ (JPY) \\
\midrule
\multicolumn{7}{@{}l}{\textit{State of emergency (17.7\%)}}\\
actual, net & 81.4 & 98,614 & 0.39 & 0.79 & 1.57 & 57,896 \\
actual, gross & 81.4 & 151,502 & 0.26 & 0.51 & 1.02 & 47,318 \\
access first ($\lambda = 0$) & 81.5 & 172,757 & 0.22 & 0.45 & 0.90 & 43,148 \\
average-oriented ($\lambda = 10$) & 81.5 & 151,147 & 0.26 & 0.51 & 1.03 & 47,448 \\
equity-constrained & 82.4 & 169,396 & 0.23 & 0.46 & 0.93 & 44,675 \\
\multicolumn{7}{@{}l}{\textit{Summer (12.5\%)}}\\
actual, net & 57.7 & 184,781 & 0.15 & 0.30 & 0.60 & 18,079 \\
actual, gross & 57.7 & 280,634 & 0.10 & 0.20 & 0.39 & $-$1,092 \\
access first ($\lambda = 0$) & 58.2 & 250,084 & 0.11 & 0.22 & 0.44 & 5,424 \\
average-oriented ($\lambda = 10$) & 57.8 & 195,190 & 0.14 & 0.28 & 0.56 & 16,024 \\
equity-constrained & 57.8 & 205,831 & 0.13 & 0.27 & 0.54 & 13,941 \\
\bottomrule
\end{tabular}
\end{table}

\paragraph{Passenger time.} Valuing the waiting of passengers who continue to travel at $v = \kappa c_h$ lowers the net saving before any loss of benefits from abandoned trips. At $\rho = 0.5$ the net change for the actual timetable falls from JPY 77,619 ($\kappa = 0$) to 66,297 ($\kappa = 0.5$) and 54,976 ($\kappa = 1$) per weekday morning in the state of emergency, and from 55,035 to 30,423 and 5,812 in the summer. In the summer little remains at $\kappa = 1$: the break-even benefit per abandoned trip (retention $\theta = 0.8$) is JPY 27 to 41, or 10 to 15\% of the mean fare, against 480 to 737 in the state of emergency. The average-oriented plan does somewhat better in the summer (net 9,598 at $\kappa = 1$, break-even 63), and the access-first plan does not survive: its long waits make the net change negative at $\kappa = 1$ ($-$9,154 in the state of emergency and $-$87,608 in the summer). The values of $\kappa$ are scenarios and not estimates; fares cancel between passengers and the operator and are not counted.

\subsection{Passenger groups}\label{sec:res-groups}

The raw smart-card transactions record the number of adults, adults with disability fares, children and children with disability fares for each tap. They carry no age field, so the groups are fare categories and not age groups. The raw counts reconcile with the cleaned taps used in the rest of the paper: over the 21 July weekdays and the same hours, the ratio of persons is 1.000 and that of records 1.000, and the per-stop correlation is 0.999. In July, adults account for 95.0\% of persons, passengers with disability fares for 2.5\% and children for 2.5\%. In the reference boardings of the two windows, adults account for 97.1\% and 96.5\%, passengers with disability fares for 2.4\% and 2.2\%, and children for 0.5\% and 1.3\% (about 9 and 59 boardings per weekday morning), so the estimates for children rest on few boardings.

Table~\ref{tab:groups} gives the added wait per boarding of each group under the actual timetable and under alternatives with different weights. Under the actual timetables the added wait per boarding is higher for children than for adults, 1.18 against 0.37 minutes in the state of emergency and 2.21 against 0.32 in the summer. Passengers with disability fares bear slightly more than adults in the state of emergency (0.58 against 0.37 minutes) but not in the summer (0.27 against 0.32). Plans that count the added wait of children three times as heavily select almost the same trips as the equal-weight plan (99\% of bus-hours in common in both windows). The added wait of children per boarding stays at 0.89 minutes in the state of emergency and moves from 0.30 to 0.29 minutes in the summer. For passengers with disability fares the corresponding changes are 0.80 to 0.78 and 0.38 to 0.39 minutes (93\% and 98\% of bus-hours in common), and the added wait of adults changes by at most 0.02 minutes. A weight of three therefore changes the plans little, because these passengers board at the same stops as the other passengers, and in these data a group weight is not a strong lever.

The older-resident proxy assigns the 239,788 residents aged 65 and over, out of the 914,219 residents assigned to a served stop (26\%), to their best stop in the baseline. It is an area-based assessment and not a count of older passengers. The added wait averages 1.98 and 1.37 minutes under the actual timetables, which is 12\% and 9\% above the resident-weighted mean (1.76 and 1.26 minutes), and 0.96 and 0.58 minutes under the average-oriented plans, close to the resident-weighted mean (0.97 and 0.59).

\begin{table}[!t]
\centering
\caption{Added expected wait per boarding (minutes) by passenger group under the actual timetable and under alternatives that weight the burden of one group by $\omega = 3$. Groups are the IC fare categories; reference boardings are the same for all plans. Disability fares combine adults and children. Older-resident areas: added wait per resident aged 65 and over, an area-based proxy (not a per-boarding value and not a count of older passengers). Plans are at the size of the reduction of each window.}
\label{tab:groups}
\small
\begin{tabular}{lrrrrr}
\toprule
Plan & Adults & Children & Disability fares & Older-resident areas & $L_1 + L_2 + L_4$ \\
\midrule
\multicolumn{6}{l}{\textit{State of emergency}}\\
actual & 0.37 & 1.18 & 0.58 & 1.98 & 52,856 \\
equal weights ($\lambda = 10$) & 0.54 & 0.89 & 0.80 & 0.96 & 534 \\
children $\omega = 3$ & 0.55 & 0.89 & 0.81 & 0.96 & 534 \\
disability fares $\omega = 3$ & 0.55 & 0.92 & 0.78 & 0.99 & 553 \\
\multicolumn{6}{l}{\textit{Summer}}\\
actual & 0.32 & 2.21 & 0.27 & 1.37 & 33,831 \\
equal weights ($\lambda = 10$) & 0.31 & 0.30 & 0.38 & 0.58 & 388 \\
children $\omega = 3$ & 0.32 & 0.29 & 0.39 & 0.58 & 388 \\
disability fares $\omega = 3$ & 0.33 & 0.30 & 0.39 & 0.60 & 392 \\
\bottomrule
\end{tabular}
\end{table}

\subsection{Topological audit of the plans}\label{sec:res-topoaudit}

Table~\ref{tab:topoaudit} lists, for each plan at the size of the actual reduction, the number of enclosed deserts, the residents inside closed-ring pockets, and the bottleneck and $W_1$ distances to the baseline diagram at one and four buses an hour. The plans of the sweep leave the topology of access almost as it was. At one bus an hour the access-first, average-oriented and equity-constrained plans keep the 18 closed rings and their 19,863 residents in both windows; the only change is one additional pocket under the state-of-emergency access-first plan (36 against 35, bottleneck distance 170~m). At four buses an hour the number of residents in closed rings changes by at most 0.32\%. The operator's reductions differ. The state-of-emergency timetable raises the closed rings from 18 to 20 and the residents inside them from 19,863 to 21,799 (+9.7\%) at one bus an hour, and from 25,602 to 28,184 (+10.1\%) at four buses an hour; the summer timetable raises the residents inside closed rings to 21,677 (+9.1\%) and 26,434 (+3.2\%). Their bottleneck distances to the baseline are 200~m and 201~m at one bus an hour and 437~m at four, and their $W_1$ distances exceed those of the access-first, average-oriented and equity-constrained plans of the same size. The counts in this section come from the stop-rate model that drives the sweep, in which the baseline has 18 closed rings with 19,863 residents. They differ from the 17 rings and 18,446 residents of the alpha complex on the located stops (Section~\ref{sec:res-deserts}), where the rings are defined on a different stop set, and slightly from the grid model of Section~\ref{sec:res-null} (19,990 residents in closed rings in the baseline against 19,863 here, and 21,799 under the summer timetable against 21,677); the differences from the grid model are below 1\% and do not alter the comparison. The bottleneck distance at one bus an hour is 194~m in the grid model and 200--201~m here. Because the sweep plans do not differ in the residents inside closed rings, the tie-break of Section~\ref{sec:tda-role} cannot select a plan different from the one that $L_F$ and waiting select. Topology is therefore a diagnostic here and not a selection criterion: it confirms that the operator's reductions enlarged closed rings that the alternatives of the same size did not.

\begin{table}[!t]
\centering
\caption{Topology of the access map under each plan at the size of the actual reduction, at one and four buses an hour ($F$, the number of buses per hour that defines service): alpha-complex pockets deeper than walking distance, closed-ring pockets and the residents in them, and bottleneck and $W_1$ distances to the baseline diagram. Plans: access first ($\lambda = 0$), average-oriented ($\lambda = 10$) and equity-constrained ($\Delta \bar w \leq 10$ with $\lambda = 10$). Plans are at the size of the reduction of each window (17.7\% and 12.5\% of trip-unit bus-hours; 22.6\% and 15.0\% of moving bus-hours in Table~\ref{tab:null}).}
\label{tab:topoaudit}
\small
\setlength{\tabcolsep}{4pt}
\begin{tabular}{@{}lrrrrrrrrr@{}}
\toprule
 & \multicolumn{5}{c}{$F = 1$ bus/h} & \multicolumn{3}{c}{$F = 4$ buses/h} & \\
\cmidrule(lr){2-6}\cmidrule(lr){7-9}
Plan & \makecell[r]{Pock-\\ets} & Rings & \makecell[r]{Resid.\\in rings} & \makecell[r]{Bottl.\\(m)} & \makecell[r]{$W_1$\\(km)} & \makecell[r]{Resid.\\in rings} & \makecell[r]{Bottl.\\(m)} & \makecell[r]{$W_1$\\(km)} & \makecell[r]{$L_1 + L_2$\\${}+ L_4$} \\
\midrule
\multicolumn{10}{@{}l}{\textit{State of emergency}}\\
baseline & 35 & 18 & 19,863 & 0 & 0.00 & 25,602 & 0 & 0.00 & 0 \\
actual & 34 & 20 & 21,799 & 200 & 1.33 & 28,184 & 437 & 1.43 & 52,856 \\
$\lambda = 0$ & 36 & 18 & 19,863 & 170 & 0.35 & 25,602 & 0 & 0.00 & 0 \\
$\lambda = 10$ & 35 & 18 & 19,863 & 0 & 0.00 & 25,683 & 113 & 0.12 & 534 \\
$\Delta \bar w \leq 10$ & 35 & 18 & 19,863 & 0 & 0.00 & 25,623 & 113 & 0.12 & 521 \\
\multicolumn{10}{@{}l}{\textit{Summer}}\\
baseline & 35 & 18 & 19,863 & 0 & 0.00 & 25,602 & 0 & 0.00 & 0 \\
actual & 34 & 19 & 21,677 & 201 & 0.83 & 26,434 & 437 & 2.17 & 33,831 \\
$\lambda = 0$ & 35 & 18 & 19,863 & 0 & 0.00 & 25,602 & 0 & 0.00 & 0 \\
$\lambda = 10$ & 35 & 18 & 19,863 & 0 & 0.00 & 25,605 & 0 & 0.00 & 388 \\
$\Delta \bar w \leq 10$ & 35 & 18 & 19,863 & 0 & 0.00 & 25,623 & 113 & 0.12 & 371 \\
\bottomrule
\end{tabular}
\end{table}

\subsection{Sensitivity analysis}\label{sec:res-sens}

Table~\ref{tab:sens} shows the added waiting per boarding when the cap is changed from 60 to 30, 120 and 240 minutes and when the directional share $\varphi$ is 0.5, 0.6 and 0.75 (the values for 0.6 lie between those for 0.5 and 0.75). The plans are held fixed. A higher cap increases the added wait of the actual timetables, from 0.34 to 0.43 minutes per boarding in the state of emergency and from 0.30 to 0.36 in the summer as the cap rises from 30 to 240 minutes, because the actual reductions leave more stops with infrequent service; the average-oriented plans change by less (0.54 to 0.55 and 0.31 to 0.33). A directional share of 0.75 lowers every value by roughly 30\% (for the state-of-emergency timetable, 0.38 to 0.27 minutes at a cap of 60) and does not change the ordering. The verdicts of Section~\ref{sec:res-pareto} are preserved. The state-of-emergency timetable has the lowest added wait of all swept plans at every cap and share tested. In the summer, the swept alternative with $\lambda = 300$ stays below the actual timetable at every cap and share (0.28 against 0.30 minutes at a cap of 30, and 0.28 against 0.34 at 60), but the lead of the $\lambda = 10$ plan over the actual timetable (0.31 against 0.34 at a cap of 60) is gone at a cap of 30 (0.31 against 0.30).

\paragraph{Capacity and card share.} The tables of Sections~\ref{sec:res-equity} to~\ref{sec:res-sens} use $(C, \sigma) = (60, 0.75)$, the setting that allows more boardings per run (Section~\ref{sec:alternatives}). All analyses were repeated with $(C, \sigma) = (50, 0.85)$. The added wait per boarding of the actual timetable and of the average-oriented and constrained plans changes by at most 0.01 minutes, the share of residents at stops with an increase of 10 minutes or more by at most 0.08 percentage points, and the largest removable shares of Table~\ref{tab:equity} by at most 0.3 percentage points. The break-even abandonment share of the actual timetable at $\rho = 0.5$ becomes 0.58 to 0.89 in the state of emergency and 0.22 to 0.34 in the summer (against 0.51 to 0.79 and 0.20 to 0.30), and no plan of Table~\ref{tab:finance} reaches $a^* = 1$ at $\rho = 0.5$ in either setting. The conclusions are the same under both settings.

\begin{table}[!t]
\centering
\caption{Added expected wait per boarding (minutes) under other values of the cap $w_{\max}$ and the directional share $\varphi$ ($\varphi = 0.5$ and $w_{\max} = 60$ is Eq.~\ref{eq:access}). The plans are the actual timetable and the swept alternatives of the same size (Section~\ref{sec:res-pareto}), chosen under $\varphi = 0.5$ and $w_{\max} = 60$ and held fixed; boardings and residents are held fixed.}
\label{tab:sens}
\small
\begin{tabular}{llrrrr}
\toprule
Plan & $\varphi$ & $w_{\max} = 30$ & $60$ & $120$ & $240$ \\
\midrule
\multicolumn{6}{l}{\textit{State of emergency}}\\
actual & 0.5 & 0.34 & 0.38 & 0.42 & 0.43 \\
 & 0.75 & 0.24 & 0.27 & 0.28 & 0.29 \\
average-oriented ($\lambda = 10$) & 0.5 & 0.54 & 0.55 & 0.55 & 0.55 \\
 & 0.75 & 0.36 & 0.37 & 0.37 & 0.37 \\
swept, $\lambda = 300$ & 0.5 & 0.48 & 0.49 & 0.53 & 0.62 \\
 & 0.75 & 0.33 & 0.34 & 0.39 & 0.46 \\
\multicolumn{6}{l}{\textit{Summer}}\\
actual & 0.5 & 0.30 & 0.34 & 0.36 & 0.36 \\
 & 0.75 & 0.22 & 0.23 & 0.24 & 0.24 \\
average-oriented ($\lambda = 10$) & 0.5 & 0.31 & 0.31 & 0.32 & 0.33 \\
 & 0.75 & 0.21 & 0.21 & 0.22 & 0.22 \\
swept, $\lambda = 300$ & 0.5 & 0.28 & 0.28 & 0.29 & 0.31 \\
 & 0.75 & 0.19 & 0.19 & 0.21 & 0.21 \\
\bottomrule
\end{tabular}
\end{table}

\section{Discussion}\label{sec:discussion}

\subsection{Comparison of the two reductions}\label{sec:disc-reductions}

Measured against chance, both 2020 reductions by Kanto Railway kept service for more residents than any random reduction of the same size at two and four buses an hour. Measured against the swept alternatives, however, the two reductions differ. In the state of emergency, when morning ridership was at a third of its usual level, the operator gave up hourly service at sparsely populated stops in order to keep waits short for the passengers who were still traveling, and no swept alternative improved on this choice in both respects. The summer timetable was prepared with time to plan, and it is repeated every year. It withdrew hourly service from stops that had almost no morning riders but as many residents as the stops it kept. A swept alternative of the same size, drawn from surplus frequency on busier corridors, would have kept hourly service for every resident at the same added waiting per boarding, and other swept alternatives would have reduced the added waiting while still leaving fewer residents without service. The two timetables thinned many of the same stops, yet only the summer timetable was dominated by the swept alternatives.

\subsection{Implications for planning}\label{sec:disc-planning}

The inputs are records that operators already collect, namely vehicle GPS and smart-card transactions, and no timetable feed is required. Reductions of the size studied here are now routine. The December 2023 reduction by Kanto Railway removed 8.5\% of weekday trips \citep{newstsukuba2023cuts}, and 70.3\% of the municipalities in Ibaraki saw reductions between 2021 and 2024 \citep{mlitkanto2025driver}. Each proposed timetable can be placed on the same frontier before it is published, together with the alternatives of the same size. The operator and the municipalities it serves can then see how many residents each option leaves without service, and at what cost in waiting. The desert map provides a second use. Because an enclosed desert is surrounded by service, a short link can reach it. On the lower-bound map, a single 1.2~km spur would bring 4,510 residents within walking distance of a stop, and 17.5~km of new route spread over ten deserts would bring 14,775.

\subsection{Contribution of topology}\label{sec:disc-topology}

The objective of the frontier (Eq.~\eqref{eq:objective}) counts residents above and below thresholds and adds the waiting of riders, and it contains no topological term, so topology does not generate the candidate plans. It enters in the description of the baseline, in the comparison of access maps and in the evaluation of every plan (Section~\ref{sec:tda-role}), and it shows three things that the conventional indicators do not. First, it separates enclosed gaps from peripheral ones and measures how deep they are, which determines whether a crossing link or an extension would close them (Section~\ref{sec:res-deserts}); two areas with the same number of residents beyond walking distance can need different remedies. Second, it provides a stable distance between access maps. This distance shows that the daily cycle reshapes access more than two years of network change, and that between the morning peak and midday the nine closed rings of service are reduced to one while the number of residents in access deserts rises by 15\%, a change that a count of the people in the gaps does not register (Section~\ref{sec:res-dynamics}). Third, it distinguishes reductions that preserve the structure of access from reductions that alter it. The actual reductions moved the one-bus diagram less than nearly all of the 6,000 random reductions, but they enlarged the closed-ring deserts by 9\%, which no random trip or duty reduction did, whereas an access-first reduction of the same size without the capacity constraint left the diagram unchanged. A count of residents losing access does not distinguish a reduction that thins the periphery from one that opens a ring in the interior. In the audit of Section~\ref{sec:res-topoaudit}, the swept plans that meet the operator's budget leave the closed rings of the baseline essentially unchanged, so among them topology does not separate plans that $L_F$ and waiting do not. It does separate the operator's reductions, which enlarged the closed rings, from all of them. Topology is therefore a diagnostic of reductions here and not a criterion that selects among the swept plans.

\subsection{Equity and financial balance}\label{sec:disc-equity}

Three results bear on equity. First, averages hide a concentration of added wait: under the actual timetables 5.9\% and 3.9\% of residents are assigned to stops whose expected wait rises by 10 minutes or more, against 0.5\% and 0.3\% under the average-oriented plans of the same size (Table~\ref{tab:equity}). Second, a minimum service standard can be met at the size of the operator’s reduction at a small cost. Capping the increase at 10 minutes removes the tail and raises the mean wait per boarding by 0.09 minutes in the state of emergency and by 0.02 minutes in the summer, and standards as strict as 5 minutes remain feasible up to about 20\% of bus-hours. The standard changes the choice of trips only in part, since 78\% and 81\% of the removed bus-hours coincide with those of the unconstrained plan. Third, giving children or passengers with disability fares three times the weight changes the plan little (99\% of bus-hours in common in both windows for children, and 93\% and 98\% for disability fares), because these passengers board at the same stops as the others. The added wait of children is high under the actual summer timetable (2.21 against 0.32 minutes for adults), but it rests on few boardings. The older-resident proxy is area-based; it shows an added wait about 12\% above the resident average under the state-of-emergency timetable and about 9\% under the summer timetable, and no difference under the average-oriented plans.

On the financial side, if half of the removed hours become paid hours saved, no plan of Table~\ref{tab:finance} saves more in labor than the fare revenue on its removed trips ($a^*$ at most 0.79), so each plan improves profit only if enough passengers keep paying. In the state of emergency the actual timetable put at most about as much revenue at risk as the alternatives of the same size ($a^*$ of 0.51 to 0.79 at $\rho = 0.5$, against 0.45 to 0.51), and in the summer about as much as the average-oriented and equity-constrained plans or more ($a^*$ of 0.20 to 0.30, against 0.22 to 0.28). The equity-constrained plan meets the 10-minute standard and still improves profit at $\rho = 0.5$ and $\theta = 0.8$ (JPY 44,675 in the state of emergency and 13,941 in the summer per weekday morning, against 47,318 to 57,896 and $-$1,092 to 18,079 for the actual timetable). These results illustrate the calculation with a national wage benchmark and with scenarios for the conversion of hours ($\rho$) and the retention of passengers ($\theta$). They are not an estimate of the operator's accounts, and they cover the weekday morning peak only.

\subsection{Relation to earlier work}\label{sec:disc-earlier}

The access-time filtration applies the waiting-time offset of \citet{hickok2024} to bus service, with waiting times derived from observed vehicle movements rather than from survey estimates. It also links the topological measurement to a frontier of operational alternatives, which earlier coverage studies did not construct. Our earlier model of Tsukuba chose bus frequency and vehicle size from travel-survey demand under an assumed level of service \citep{nakamura2025parkandride}. The measured service used here can replace those assumptions, and the frontier adds the coverage side of the choice.

\subsection{Limitations and future work}\label{sec:scope}

Service is measured in two ways that bracket what was offered. The footprint definition counts all moving vehicles on roads used by revenue trips, and the card-tap definition counts only the trips on which passengers were recorded. The desert map holds under both definitions. Waiting follows the half-headway convention, which is the standard assumption for frequent service, and a model of timed arrivals on infrequent routes \citep{fan2009wait,ingvardson2018arrival} would be a natural refinement. The waits are expected waits for randomly arriving passengers in a model with equal frequencies in the two directions and a cap of 60 minutes (Section~\ref{sec:topology}), and the distributions in Section~\ref{sec:res-equity} are spatial distributions of these expected waits, not of waits that individual passengers realize. Walking access is measured as straight-line distance at a uniform speed, not along the street network. This can overstate access where a river, a railway or a limited crossing separates a resident from a nearby stop. We checked this for the five largest closed-ring pockets, which contain 76\% of the closed-ring population, by routing from each deepest point to the nearest served stop on the OpenStreetMap street network. The network distance was 1.14--1.29 times the straight-line distance (mean 1.20), and each pocket remained beyond 800~m. The straight-line assumption therefore does not create these enclosures. Smaller or more irregular pockets have not been checked individually, and until they are, they should be read as indicative rather than exact.

A capacity-feasible alternative is not necessarily an operationally schedulable one. The frontier is traced with route pairs in 30-minute blocks and with 72\% of bus-hours eligible for removal, and it is checked only against the average boardings per remaining run on each route pair. It does not represent driver duties, vehicle blocks and deadheading, minimum turnaround, relief points or transfer connectivity. A swept alternative that saves the target bus-hours on paper may therefore not correspond to a schedule that a planner could operate, and the average load per run can mask an overloaded individual run. The frontier is best read as a strategic screening tool rather than as a timetable generator. It shows which trade-offs between access and waiting can be reached in principle, and it narrows the search for a planner, who then checks candidate schedules against these constraints. A planner who works trip by trip, or who can re-optimize duties and blocks rather than only routes, has more options than the frontier we trace. In terms of access, the frontier is therefore a conservative estimate of what can be achieved, even though it is not a certified schedule. The frontier itself is traced by a lazy greedy sweep over ten values of a scalarization weight, and it is not solved exactly. The two verdicts above therefore mean that no alternative found by this search dominates the actual timetable, and not that no such alternative exists. An exact or more exhaustively searched frontier could in principle find one, particularly for the state-of-emergency timetable, for which no dominating alternative was found. The analysis covers one operator and two reductions in 2020, and our GPS data end in June 2022. To apply the method to another operator or period, the same inputs are needed: vehicle GPS and smart-card records with recoverable route and direction labels, and population data at a comparable resolution. We expect the method to apply with little change to further reductions by this operator, and with more preparation elsewhere.

\section{Conclusion}

Persistent homology of walking and waiting time, computed from GPS and smart-card records without a timetable, locates 35 enclosed service deserts around Tsukuba. Seventeen of them are fully ringed by walkable service and are home to 18,446 residents. The same analysis shows that the access map changes more between the morning peak and midday than over two years. Applied to two timetable reductions made by Kanto Railway in 2020, the framework shows that both protected service at two and four buses an hour better than any random reduction of the same size. No swept alternative improved on the state-of-emergency timetable in both access and waiting. The summer school-holiday timetable was different. It withdrew hourly service from 5,253 residents around stops with almost no morning riders, while swept alternatives of the same size would have kept hourly service for all of them at the same added waiting, or would have reduced the added waiting while still serving more residents. These alternatives are heuristic and capacity-feasible, but they are not certified as optimal and they are not fully schedulable, so the sweep is best read as a screening step ahead of detailed timetable design.

Averages also hide who bears a reduction. Under the actual timetables 5.9\% and 3.9\% of residents are assigned to stops whose expected wait rises by 10 minutes or more, against 0.5\% and 0.3\% in swept plans of the same size, and a cap of 10 minutes on the increase can be met at the operator's size at a small cost in average waiting. At a benchmark wage and with half of the removed hours converted to paid hours saved, the labor saving of the operator's timetables and of the main alternatives falls short of the fare revenue on their removed trips, so the gain depends on how many passengers keep paying; a plan that meets a 10-minute standard improves profit in both windows if 80\% of that revenue is retained. As driver shortages force further reductions, operators can use data they already hold to narrow the search for reductions that save the same bus-hours while keeping more residents within reach of a bus.

\section*{Acknowledgments}
The authors gratefully acknowledge Kanto Railway for providing the bus GPS and IC data that made this research possible.
This work was supported by JSPS KAKENHI, Grant Numbers JP25K24393 and JP26K14726.
This work was also supported by JST ERATO, Grant Number JPMJER2301, Japan.

\section*{Declaration of generative AI and AI-assisted technologies in the manuscript preparation process}
During the preparation of this work the authors used Claude (Anthropic) to revise the wording of the manuscript and the supplementary material, to draft and debug analysis and plotting code, and to check the text against the numerical outputs of the analysis. After using this tool, the authors reviewed and edited the content as needed and take full responsibility for the content of the published article. The figures were drawn from the data by plotting code; no generative image tool was used.

\bibliographystyle{apalike}
\bibliography{references}

\end{document}